\documentclass[aps,twocolumn,pra,superscriptaddress,showpacs,tightenlines]{revtex4}
\usepackage{amssymb}
\usepackage{amsmath}
\usepackage{graphicx}
\usepackage{epsfig}
\usepackage{subfigure}
\usepackage{amsfonts}
\usepackage{txfonts}

\begin{document}
\title{Coherent excitation-energy transfer and quantum entanglement in a dimer}
\author{Jie-Qiao Liao}
\affiliation{Institute of Theoretical Physics, Chinese Academy of
Sciences, Beijing 100190, China}
\author{Jin-Feng Huang}
\affiliation{Key Laboratory of Low-Dimensional Quantum Structures
and Quantum Control of Ministry of Education, and Department of
Physics, Hunan Normal University, Changsha 410081, China}
\affiliation{Institute of Theoretical Physics, Chinese Academy of
Sciences, Beijing 100190, China}
\author{Le-Man Kuang}
\affiliation{Key Laboratory of Low-Dimensional Quantum Structures
and Quantum Control of Ministry of Education, and Department of
Physics, Hunan Normal University, Changsha 410081, China}
\author{C. P. Sun}
\affiliation{Institute of Theoretical Physics, Chinese Academy of
Sciences, Beijing 100190, China}
\date{\today}

\begin{abstract}
We study coherent energy transfer of a single excitation and quantum
entanglement in a dimer, which consists of a donor and an acceptor
modeled by two two-level systems. Between the donor and the
acceptor, there exists a dipole-dipole interaction, which provides
the physical mechanism for coherent energy transfer and entanglement
generation. The donor and the acceptor couple to two independent
heat baths with diagonal couplings that do not dissipate the energy
of the non-coupling dimer. Special attention is paid to the effect
on single-excitation energy transfer and entanglement generation of
the energy detuning between the donor and the acceptor and the
temperatures of the two heat baths. It is found that, the
probability for single-excitation energy transfer largely depends on
the energy detuning in the low temperature limit. Concretely, the
positive and negative energy detunings can increase and decrease the
probability at steady state, respectively. In the high temperature
limit, however, the effect of the energy detuning on the probability
is negligibly small. We also find that the probability is negligibly
dependent on the bath temperature difference of the two heat baths.
In addition, it is found that quantum entanglement can be generated
in the process of coherent energy transfer. As the bath temperature
increases, the generated steady-state entanglement decreases. For a
given bath temperature, the steady-state entanglement decreases with
the increase of the absolute value of the energy detuning.
\end{abstract}
\narrowtext


\maketitle
\section{\label{Sec:1}Introduction}

Coherent excitation energy transfer is an important step of
photosynthesis~\cite{Blankenship}, in which photosynthetic pigments
capture the solar light to create electronic excitations and then
transfer the excitation energy to a reaction
center~\cite{May-Kuhn,Fleming1994,Fleming2009,Renger2009,Fleming2007,Flemingnature}.
Usually, the transfer of a single excitation from the pigment where
the electronic excitation is created to the reaction center is a
very complicated physical process, since the practical transfer
process takes place on a complicated network of pigments. However,
the basic physical mechanism can be revealed in such a
light-harvesting complex by studying a basic part: a dimer system
which consists of a donor and an acceptor modeled by two two-level
systems.

On a complicated network of pigments, there generally exist two
kinds of interactions. On one hand, between any two pigments there
exists a dipole-dipole interaction, which results in excitation
energy transfer. On the other hand, the pigments interact inevitably
with their surrounding environments such as the nuclear degrees of
freedom and the proteins. Corresponding to different cases for the
scale of the two kind of interactions, different approaches have
been proposed to study the single-excitation energy transfer.
Concretely, when the dipole-dipole interactions between any two
pigments are much weaker than the interactions of the pigments with
their environments, the energy transfer process can be well
characterized by the F\"{o}rster theory~\cite{Forster1948}, in which
the evolution of the network is calculated perturbatively up to the
second order in the dipole-dipole interactions between the pigments;
When the interactions of the pigments with their environments are
much weaker than the dipole-dipole interactions between any two
pigments, various approaches based on the quantum master equation
have been proposed (e.g.,
Refs.~\cite{Ishizaki2009,Jang2008,Palmieri2009,Aspuru-Guzik2008,Aspuru-Guzik20091,
Aspuru-Guzik20092,Aspuru-Guzik20093,Plenio2008,Plenio20091,Castro2008,Castro2009,Nazir2009,Nori2009,Liang2010,Yang2010}),
in which the evolution of the network is calculated perturbatively
up to the second order in the interactions between the pigments and
their environments.

With the above considerations, in this article we study
single-excitation energy transfer in a dimer, which consists of a
donor and an acceptor modeled by two two-level systems. Obviously,
when the donor and the acceptor are decoupled, it is impossible to
realize energy transfer between them. Therefore, the simplest way to
realize energy transfer is to turn on a non-trivial interaction (for
example, the dipole-dipole interaction) between the donor and the
acceptor. Then a single excitation can coherently oscillate between
the donor and the acceptor. However, in this case, there is no
steady-state energy transfer, namely the transferred energy can not
approach to a stationary value. In the presence of environments, the
donor and the acceptor will inevitably couple with environments. In
general, the coupling form between the donor (acceptor) and its
environment is diagonal in the representation of the free
Hamiltonian of the donor (acceptor). Physically, due to this type of
coupling, although the excitation energy will not decay into the
environments, it will induce a steady-state energy transfer between
the donor and the acceptor. Since in practical cases both the
characteristic frequency and the heat bath temperatures of the donor
and the acceptor may be different due to different chemical
structures, we study in detail how the characteristic frequencies
and the heat bath temperatures of the donor and acceptor affect the
efficiency of the excitation energy transfer. This is one point of
the motivations of our present work.

In the presence of the interactions between the pigments for
transferring energy, a naturally arising question is how about the
quantum entanglement among the pigments which are involved in the
energy transfer process. Because quantum entanglement is at the
heart of the foundation of quantum
mechanics~\cite{Bell1987,Einstein1935} and quantum information
science~(e.g., Refs.~\cite{Nielsen2000,Qian2005}), it is interesting
to know how is the dynamics of the created quantum entanglement in
the dimer system during the process of single-excitation energy
transfer. This is the other point of the motivations of our present
investigations. In fact, recently people have become aware of
quantum entanglement in some chemical and biologic systems (e.g.,
Refs.~\cite{Briegel2008,Briegel2009,Plenio20091,Thorwart2009,Sarovar2009,Caruso2009})
such as photosynthetic light-harvesting
complexes~\cite{Plenio20091,Sarovar2009,Caruso2009}.

This article is organized as follows: In Sec.~\ref{Sec:2}, we
present the physical model and the Hamiltonian for studying the
single-excitation energy transfer. A dimer consists of a donor and
an acceptor, which are immersed in two independent heat baths.
Between the donor and the acceptor, there exists a dipole-dipole
interaction, which provides the physical mechanism for coherent
excitation energy transfer and entanglement generation. In
Sec.~\ref{Sec:3}, we derive a quantum master equation to describe
the evolution of the dimer. Based on the quantum master equation we
obtain optical Bloch equations and their solutions. In
Sec.~\ref{Sec:4}, we study single-excitation energy transfer from
the donor to the acceptor. The effect on the transfer probability of
the energy detuning and the bath temperatures are studied carefully.
In Sec.~\ref{Sec:5}, we study the quantum entanglement between the
donor and the acceptor by calculating the concurrence. We conclude
this work with some remarks in Sec.~\ref{Sec:6}. Finally, we give an
appendix for derivation of quantum master
equation~(\ref{mastereqfordiagonalcase}).

\section{\label{Sec:2}Physical model and Hamiltonian}

\begin{figure}[tbp]
\includegraphics[bb=45 419 407 757, width=8 cm]{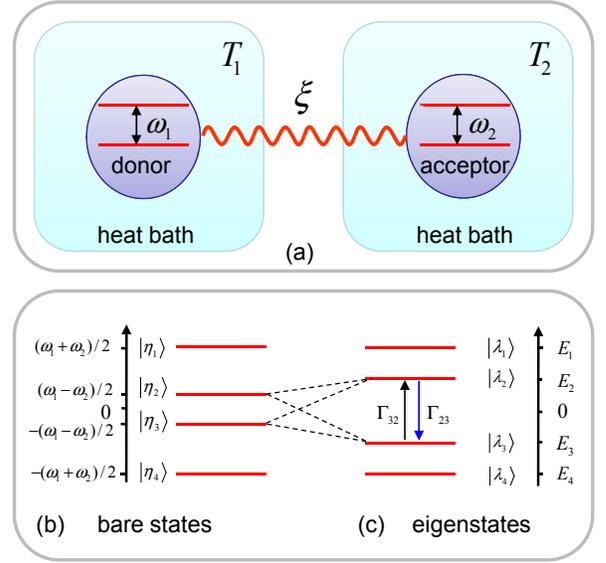}
\caption{(Color online) (a) Schematic of the physical system. A
donor and an acceptor are immersed in two independent heat baths of
temperatures $T_{1}$ and $T_{2}$, respectively. A dipole-dipole
interaction of strength $\xi$ exists between the donor and the
acceptor, which are described by two two-level systems with resonant
frequencies $\omega_{1}$ and $\omega_{2}$, respectively. (b) The
energy levels of the bare states $|\eta_{n}\rangle$ ($n=1,2,3,4$) of
the donor and the acceptor when they are decoupling. (c) The energy
levels of the eigenstates $|\lambda_{n}\rangle$ ($n=1,2,3,4$) of the
coupled donor and acceptor. The corresponding eigen-energies are
denoted by $E_{n}$. The parameters $\Gamma_{23}$ and $\Gamma_{32}$
are, respectively, the bath induced transition rates from states
$|\lambda_{2}\rangle$ to $|\lambda_{3}\rangle$ and from states
$|\lambda_{3}\rangle$ to $|\lambda_{2}\rangle$.}\label{schematic}
\end{figure}

As illustrated in Fig.~\ref{schematic}(a), the physical system under
our consideration is a dimer, which consists of a donor and an
acceptor modeled by two two-level systems (TLSs), TLS$1$ (donor) and
TLS$2$ (acceptor), with respective energy separations $\omega_{1}$
and $\omega_{2}$. The donor and the acceptor are immersed in two
independent heat baths of temperatures $T_{1}$ and $T_{2}$,
respectively. Between the donor and the acceptor there exists a
dipole-dipole interaction of strength $\xi$.
The Hamiltonian of the total system, including the two coupled TLSs
and their heat baths, is composed of three parts,
\begin{equation}
H=H_{\textrm{TLSs}}+H_{B}+H_{I},\label{Hamiltonian}
\end{equation}
where $H_{\textrm{TLSs}}$ is the Hamiltonian (with $\hbar=1$) of the
two coupled TLSs,
\begin{equation}
H_{\textrm{TLSs}}=\frac{\omega _{1}}{2}\sigma _{1}^{z}+\frac{\omega
_{2}}{2}\sigma _{2}^{z}+\xi \left( \sigma _{1}^{+}\sigma
_{2}^{-}+\sigma _{1}^{-}\sigma _{2}^{+}\right).\label{HofTLSs}
\end{equation}
Concretely, the first two terms in Eq.~(\ref{HofTLSs}) are free
Hamiltonians of the two TLSs, which are described by the usual Pauli
operators
$\sigma_{l}^{+}=(\sigma_{l}^{-})^{\dag}=(\sigma_{x}+i\sigma_{y})/2=\left\vert
e\right\rangle_{ll} \left\langle g\right\vert$ and
$\sigma_{l}^{z}=\left\vert e\right\rangle_{ll}\left\langle
e\right\vert-\left\vert g\right\rangle_{ll} \left\langle
g\right\vert$, where $|g\rangle_{l}$ and $|e\rangle_{l}$ are,
respectively, the ground and excited states of the $l$th ($l=1,2$)
TLS, namely TLS$l$. The last term in Eq.~(\ref{HofTLSs}) depicts the
dipole-dipole interaction of strength $\xi$ between the two TLSs.
This dipole-dipole interaction provides the physical mechanism for
excitation energy transfer and entanglement generation between the
two TLSs.

The Hilbert space of the donor and the acceptor is of four dimension
with the four basis states $|\eta_{1}\rangle=|ee\rangle$,
$|\eta_{2}\rangle=|eg\rangle$, $|\eta_{3}\rangle=|ge\rangle$, and
$|\eta_{4}\rangle=|gg\rangle$, as shown in Fig.~\ref{schematic}(b).
In the presence of the dipole-dipole interaction, a stationary
single-excitation state should be delocalized and composed of a
combination of the single-excitation in the two TLSs. According to
Hamiltonian~(\ref{HofTLSs}), we can obtain the following four
eigenstates
\begin{eqnarray}
\left\vert \lambda _{1}\right\rangle &=&\left\vert
ee\right\rangle,\nonumber\\
\left\vert \lambda
_{2}\right\rangle&=&\cos\left(\theta/2\right)\left\vert
eg\right\rangle +\sin\left(\theta/2\right) \left\vert
ge\right\rangle,\nonumber\\
\left\vert \lambda
_{3}\right\rangle&=&-\sin\left(\theta/2\right)\left\vert
eg\right\rangle +\cos\left(\theta/2\right)\left\vert
ge\right\rangle,\nonumber\\
\left\vert \lambda
_{4}\right\rangle&=&\left\vert gg\right\rangle,\label{eigenstates}
\end{eqnarray}
and the corresponding eigenenergies
$E_{1}=-E_{4}=(\omega_{1}+\omega_{2})/2$ and
$E_{2}=-E_{3}=\sqrt{\Delta\omega^{2}/4+\xi^{2}}$, as shown in
Fig.~\ref{schematic}(c), by solving the eigen-equation
$H_{\textrm{TLSs}}\left\vert\lambda
_{n}\right\rangle=E_{n}\left\vert\lambda _{n}\right\rangle$
($n=1,2,3,4$). Here we introduce the energy detuning
$\Delta\omega=\omega_{1}-\omega_{2}$ and the mixing angle $\theta$
defined by $\tan\theta=2\xi/\Delta\omega$. Note that here the mixing
angle $0<\theta<\pi$. Therefore, when $\Delta\omega>0$, namely
$\omega_{1}>\omega_{2}$, we have
$\theta=\arctan(2\xi/\Delta\omega)$; however, when $\Delta\omega<0$,
that is $\omega_{1}<\omega_{2}$, we have
$\theta=\arctan(2\xi/\Delta\omega)+\pi$.

As pointed out by Caldeira and Leggett~\cite{Leggettnp}, when the
couplings of a system with its environment are weak, it is universal
to model the environment of the system as a harmonic oscillator heat
bath. In this work, we suppose that the couplings of the TLSs with
their environments are weak, then it is reasonable to model the
environments as two harmonic oscillator heat baths with the
Hamiltonian
\begin{eqnarray}
H_{B}&=&H^{(a)}_{B}+H^{(b)}_{B}.
\end{eqnarray}
Here $H^{(a)}_{B}$ and $H^{(b)}_{B}$ are respectively the
Hamiltonians of the heat baths for the TLS$1$ and TLS$2$,
\begin{eqnarray}
H^{(a)}_{B}&=&\sum_{j}\omega _{aj}a_{j}^{\dagger }a_{j},\hspace{0.5
cm} H^{(b)}_{B}=\sum_{k}\omega _{bk}b_{k}^{\dagger }b_{k},
\end{eqnarray}
where $a^{\dag}_{j}$ ($b^{\dag}_{k}$) and $a_{j}$ ($b_{k}$) are,
respectively, the creation and annihilation operators of the $j$th
($k$th) harmonic oscillator with frequency $\omega_{aj}$
($\omega_{bk}$) of the heat bath for TLS$1$ (TLS$2$). In practical
systems of excitation energy transfer, the environment is composed
of the nuclear degrees of freedom of the molecules.

The interaction Hamiltonian of the TLSs with their heat baths
reads~(e.g.,
Refs.~\cite{Ishizaki2009,Jang2008,Palmieri2009,Aspuru-Guzik2008,Aspuru-Guzik20091,
Aspuru-Guzik20092,Aspuru-Guzik20093,Plenio2008,Plenio20091})
\begin{equation}
H_{I}=\sigma_{1}^{+}\sigma_{1}^{-}\sum_{j}g_{1j}(a_{j}^{\dagger
}+a_{j})+\sigma_{2}^{+}\sigma_{2}^{-}\sum_{k}g_{2k}(b_{k}^{\dagger}+b_{k}).\label{diacouplingH}
\end{equation}
In this case, there is no energy exchange between the TLSs and their
heat baths. This type of diagonal coupling has been used to describe
the dephasing of quantum systems~\cite{Gao2007}. For simplicity, but
without loss of generality, in the following we assume the coupling
strengthes $g_{1j}$ and $g_{2k}$ are real numbers.

\section{\label{Sec:3}Quantum master equation and optical Bloch equations}

Generally speaking, there are two kinds of different approaches to
study photonsynthetic excitation energy transfer. One is based on
the F\"{o}rster theory~\cite{Forster1948}, which is valid when the
electronic couplings between pigments are smaller than the couplings
between electrons and environments. The other is usually based on
quantum master
equations~\cite{Ishizaki2009,Jang2008,Palmieri2009,Aspuru-Guzik2008,Aspuru-Guzik20091,
Aspuru-Guzik20092,Aspuru-Guzik20093,Plenio2008,Plenio20091,Castro2008,Castro2009,Nazir2009,Nori2009,Liang2010,Yang2010}
in various forms, which are valid when the electron-environment
couplings are weaker than electronic couplings between pigments. In
this work, we shall consider the latter case where the coupling
(with strength $\xi$) between the two TLSs is stronger than the
couplings (relating to $\gamma$) between the TLSs and their local
environments (in our following considerations we take
$\xi/\gamma=5$). We will derive a quantum master equation by
truncating the evolution up to the second order in the
TLS-environment couplings. On the other hand, we derive the master
equation in the eiegen-representation of the two coupled TLSs so we
may safely make the secular approximation~\cite{Breuer} by
neglecting the high-frequency oscillating terms. This approximation
is also equivalent to rotating wave approximation in quantum optical
systems. The detailed derivation of the quantum master equation will
be presented in the appendix.

In the eigen-representation of Hamiltonian~(\ref{HofTLSs}) of the
two coupled TLSs, the quantum master equation in Schr\"{o}dinger
picture reads,
\begin{eqnarray}
\label{mastereqfordiagonalcase} \dot{\rho}_{S}
&=&i[\rho_{S},H_{\textrm{TLSs}}]\nonumber\\
&&+\sum_{n=1,2,3}\Pi_{n}(2\sigma _{nn}\rho
_{S}\sigma_{nn}-\sigma_{nn}\rho _{S}-\rho _{S}\sigma _{nn})\nonumber\\
&&+\Gamma_{32}(2\sigma_{23}\rho_{S} \sigma_{32}-\sigma_{33}\rho_{S}-\rho_{S}\sigma_{33})\nonumber\\
&&+\Gamma_{23}(2\sigma_{32}\rho _{S}\sigma _{23}-\sigma _{22}\rho
_{S}-\rho_{S}
\sigma _{22})\nonumber\\
 &&+2X_{12}(\sigma _{11}\rho _{S}\sigma
_{22}+\sigma_{22}\rho _{S}\sigma _{11})\nonumber\\
&&+2X_{13}(\sigma _{11}\rho _{S}
\sigma_{33}+\sigma_{33}\rho _{S}\sigma _{11})\nonumber\\
&&+2X_{23}(\sigma _{33}\rho _{S}\sigma_{22}+\sigma_{22}\rho
_{S}\sigma _{33}).
\end{eqnarray}
In Eq.~(\ref{mastereqfordiagonalcase}), $\rho_{S}$ is the reduced
density matrix of the two TLSs. The transition operators
$\sigma_{nm}$ ($n,m=1$, $2$, $3$, and $4$) are defined as
$\sigma_{nm}\equiv|\lambda_{n}\rangle\langle\lambda_{m}|$, where the
states $|\lambda_{n}\rangle$ have been defined in
Eq.~(\ref{eigenstates}). Meanwhile, we introduce the effective rates
as follows:
\begin{eqnarray}
\label{decayfactors}
\Pi_{1} &=&\chi_{1}+\chi_{2},\nonumber\\
\Pi_{2} &=&\cos^{4}(\theta/2)\chi_{1}
+\sin^{4}(\theta/2)\chi_{2},\nonumber\\
\Pi _{3} &=&\sin ^{4}(\theta/2)\chi_{1}+\cos
^{4}(\theta/2)\chi_{2},\nonumber\\
\Gamma _{32}&=&\frac{1}{4}\sin ^{2}\theta[\gamma
_{1}\bar{n}_{1}(\varepsilon)+\gamma
_{2}\bar{n}_{2}(\varepsilon)],\nonumber\\
\Gamma _{23}&=&\frac{1}{4}\sin ^{2}\theta[\gamma _{1}( \bar{n}_{1}(
\varepsilon )+1)+\gamma_{2}(\bar{n}_{2}(\varepsilon)+1)],\nonumber\\
X_{12} &=&\cos^{2}(\theta/2)\chi_{1}+\sin
^{2}(\theta/2)\chi_{2},\nonumber\\
X_{13} &=&\sin ^{2}(\theta/2)\chi_{1}+\cos
^{2}(\theta/2)\chi_{2},\nonumber\\
X_{23} &=&\frac{1}{4}\sin ^{2}\theta (\chi_{1}+\chi_{2}),
\end{eqnarray}
where
$\chi_{l}=\lim_{\omega\rightarrow0}S_{l}(\omega)[2\bar{n}_{l}(\omega)+1]$,
with $S_{l}(\omega)=\pi\varrho_{l}(\omega)g_{l}^{2}(\omega)$ and
$\gamma_{l}=\pi\varrho_{l}(\varepsilon) g_{l}^{2}(\varepsilon)$ for
$l=1,2$. Here $\varrho_{1}(\omega)$ and $\varrho_{2}(\omega)$ are
respectively the densities of state for the two independent heat
baths surrounding the donor and the acceptor. The parameter
$\varepsilon\equiv E_{2}-E_{3}$ is the energy separation between the
two eigenstates $|\lambda_{2}\rangle$ and $|\lambda_{3}\rangle$. And
\begin{eqnarray}
\bar{n}_{l}(\omega)=\frac{1}{\exp(\omega/T_{l})-1}
\end{eqnarray}
is the thermal average excitation numbers of the heat baths of
TLS$l$. Hereafter we set the Boltzmann constant $k_{B}=1$. We
consider a special case of the ohmic spectrum densities
$S_{1}(\omega)=\eta_{1}\omega$ and $S_{2}(\omega)=\eta_{2}\omega$,
and then we obtain $\chi_{1}=2\eta_{1}T_{1}$ and
$\chi_{2}=2\eta_{2}T_{2}$.

From quantum master equation~(\ref{mastereqfordiagonalcase}), we can
see that there exist both dissipation and dephasing processes in the
eigen-representation of the Hamiltonian~(\ref{HofTLSs}). The first
line in Eq.~(\ref{mastereqfordiagonalcase}) describes the unitary
evolution of the system under the Hamiltonian ~(\ref{HofTLSs}). The
second line in Eq.~(\ref{mastereqfordiagonalcase}) describes the
dephasing of the states $|\lambda_{1}\rangle$,
$|\lambda_{2}\rangle$, and $|\lambda_{3}\rangle$. The third and
fourth lines describe, respectively, the exciting process from
$|\lambda_{3}\rangle$ to $|\lambda_{2}\rangle$ and the decay process
from $|\lambda_{2}\rangle$ to $|\lambda_{3}\rangle$, as illustrated
in Fig.~\ref{schematic}(b). Moreover, there exist three cross
dephasing processes in the last three lines in
Eq.~(\ref{mastereqfordiagonalcase}), these terms can decrease the
coherence between two levels, which can be seen from the following
optical Bloch equations~(\ref{OBEfordiagonal}).

According to quantum master
equation~(\ref{mastereqfordiagonalcase}), we can derive optical
Bloch equations for the elements
$\langle\sigma_{mn}(t)\rangle=\textrm{Tr}_{S}[\rho_{s}(t)\sigma_{mn}]$,
\begin{eqnarray}
\label{OBEfordiagonal}
\left\langle \dot{\sigma}_{11}\left( t\right) \right\rangle&=&\left\langle \dot{\sigma}_{44}\left( t\right) \right\rangle=0,\nonumber\\
\left\langle \dot{\sigma}_{22}\left( t\right) \right\rangle
&=&-\left\langle \dot{\sigma}_{33}\left( t\right)
\right\rangle=2\Gamma _{32}\left\langle \sigma _{33}\left( t\right)
\right\rangle -2\Gamma
_{23}\left\langle \sigma _{22}\left( t\right) \right\rangle,\nonumber\\
\left\langle \dot{\sigma}_{32}\left( t\right) \right\rangle
&=&[-i\varepsilon-\left( \Pi _{2}+\Pi _{3}+\Gamma _{23}+\Gamma
_{32}-2X_{23}\right)] \left\langle \sigma _{32}\left( t\right)
\right\rangle.\nonumber\\
\end{eqnarray}
Here we present only the equations of motion for the elements which
will be used below. In fact, the equations of motion for all of the
elements in the density matrix $\rho_{S}$ can be obtained according
to quantum master equation~(\ref{mastereqfordiagonalcase}). Clearly,
from optical Bloch equations~(\ref{OBEfordiagonal}) we can see that
the diagonal elements decouple with the off-diagonal elements. It is
straightforward to get the transient solutions of optical Bloch
equations~(\ref{OBEfordiagonal}),
\begin{eqnarray}
\label{transientsolution2} \left\langle \sigma _{11}\left( t\right)
\right\rangle&=&\left\langle \sigma _{11}\left( 0\right)
\right\rangle,\hspace{0.5 cm}\left\langle \sigma _{44}\left(
t\right) \right\rangle=\left\langle
\sigma _{44}\left( 0\right) \right\rangle,\nonumber\\
\left\langle \sigma _{22}\left( t\right) \right\rangle
&=&\frac{\left( \left\langle \sigma _{22}\left( 0\right)
\right\rangle +\left\langle \sigma _{33}\left( 0\right)
\right\rangle \right) \Gamma _{32}}{\Gamma _{23}+\Gamma
_{32}}\nonumber\\
&&+\frac{\left( \left\langle \sigma _{22}\left( 0\right)
\right\rangle \Gamma _{23}-\left\langle \sigma _{33}\left( 0\right)
\right\rangle \Gamma _{32}\right) }{\Gamma _{23}+\Gamma
_{32}}e^{-2\left( \Gamma _{23}+\Gamma
_{32}\right) t},\nonumber\\
\left\langle \sigma _{33}\left( t\right) \right\rangle
&=&\frac{\left( \left\langle \sigma _{22}\left( 0\right)
\right\rangle +\left\langle \sigma _{33}\left( 0\right)
\right\rangle \right) \Gamma _{23}}{\Gamma _{23}+\Gamma
_{32}}\nonumber\\
&&+\frac{\left( \left\langle \sigma _{33}\left( 0\right)
\right\rangle \Gamma _{32}-\left\langle \sigma _{22}\left( 0\right)
\right\rangle \Gamma _{23}\right) }{\Gamma _{23}+\Gamma
_{32}}e^{-2\left( \Gamma _{23}+\Gamma
_{32}\right) t},\nonumber\\
\left\langle \sigma _{32}\left( t\right) \right\rangle
&=&\left\langle \sigma _{32}\left( 0\right) \right\rangle e^{-\left(
\Gamma _{23}+\Gamma _{32}+\cos^{2}\theta\Pi_{1}\right)
t}e^{-i\varepsilon t}.
\end{eqnarray}
Here we have used the relation $\Pi _{2}+\Pi
_{3}-2X_{23}=\cos^{2}\theta\Pi_{1}$. The steady-state solutions of
Eq.~(\ref{transientsolution2}) read
\begin{eqnarray}
\left\langle \sigma _{11}\left(\infty\right) \right\rangle
&=&\left\langle \sigma _{11}\left( 0\right)
\right\rangle,\hspace{0.5 cm}\left\langle \sigma _{44}\left(
\infty\right) \right\rangle=\left\langle
\sigma _{44}\left( 0\right) \right\rangle,\nonumber\\
\left\langle \sigma _{22}\left(\infty\right) \right\rangle
&=&\frac{\left( \left\langle \sigma _{22}\left( 0\right)
\right\rangle +\left\langle \sigma _{33}\left( 0\right)
\right\rangle \right) \Gamma _{32}}{\Gamma _{23}+\Gamma
_{32}},\nonumber\\
\left\langle \sigma _{33}\left(\infty\right) \right\rangle
&=&\frac{\left( \left\langle \sigma _{22}\left( 0\right)
\right\rangle +\left\langle \sigma _{33}\left( 0\right)
\right\rangle \right) \Gamma _{23}}{\Gamma _{23}+\Gamma
_{32}},\nonumber\\
\left\langle \sigma _{32}\left(\infty\right)
\right\rangle&=&0.\label{steastate}
\end{eqnarray}
The steady-state solutions for other off-diagonal elements of the
density matrix are zero. Therefore, we can see that the steady state
of the two TLSs is a completely mixed one.

\section{\label{Sec:4}Probability for single-excitation energy transfer}

In order to study the probability for single-excitation energy
transfer from the TLS$1$ (donor) to the TLS$2$ (acceptor), we assume
that the TLS$1$ initially possesses a single excitation and the
TLS$2$ is in its ground state, which means the initial state of the two
TLSs is
\begin{eqnarray}
\left\vert\varphi\left(0\right)\right\rangle _{S}&=&\left\vert
eg\right\rangle=\cos\left(\theta/2\right) \left\vert
\lambda_{2}\right\rangle-\sin \left(\theta/2\right) \left\vert
\lambda _{3}\right\rangle.\label{initialstate2}
\end{eqnarray}
Since the couplings between the TLSs and their heat baths are
diagonal, there is no energy exchange between the TLSs and their
heat baths, and the probability for finding the TLS$2$ in its
excited state is right that of the single excitation energy
transfer,
\begin{eqnarray}
P(t)&\equiv&\textrm{Tr}_{2}[\rho_{2}\sigma^{+}_{2}\sigma^{-}_{2}]\nonumber\\
&=&\langle\sigma_{11}(t)\rangle+\sin^{2}(\theta/2)\langle\sigma_{22}(t)\rangle+\cos^{2}(\theta/2)\langle\sigma_{33}(t)\rangle\nonumber\\
&&+\sin\theta
\textrm{Re}[\langle\sigma_{23}(t)\rangle],\label{probabilityformula}
\end{eqnarray}
where $\rho_{2}=\textrm{Tr}_{1}[\rho_{S}]$ is the reduced density
matrix of the TLS$2$.

\subsection{Transient state case}

According to Eq.~(\ref{transientsolution2}), the probability given
in Eq.~(\ref{probabilityformula}) can be expressed as follows:
\begin{eqnarray}
P\left( t\right) &=&\frac{\Gamma _{32}\sin^{2}(\theta/2)+\Gamma
_{23}\cos^{2}(\theta/2)}
{\Gamma _{23}+\Gamma _{32}}\nonumber\\
&&+\cos\theta\frac{\Gamma _{32}\sin^{2}(\theta/2)-\Gamma
_{23}\cos^{2}(\theta/2)} {\Gamma _{23}+\Gamma _{32}}e^{-2\left(
\Gamma _{23}+\Gamma _{32}\right) t}\nonumber\\
&&-\frac{1}{2}\sin ^{2}\theta \cos(\varepsilon t)e^{-\left(\Gamma
_{23}+\Gamma
_{32}+\cos^{2}\theta\Pi_{1}\right)t}.\label{probability2}
\end{eqnarray}
Now, we obtain the probability for single-excitation energy transfer
from the TLS$1$ to TLS$2$. This probability~(\ref{probability2}) is
a complicated function of the variables of the two TLSs and their
heat baths, such as the energy separations $\omega_{1}$ and
$\omega_{2}$, the strength $\xi$ of the dipole-dipole interaction,
and the temperatures $T_{1}$ and $T_{2}$ of the heat baths. To see
clearly the effect on probability~(\ref{probability2}) of the bath
temperatures and the energy separations of the TLSs, we introduce
the following variables: mean temperature $T_{m}=(T_{1}+T_{2})/2$,
mean energy separation $\omega _{m}=(\omega _{1}+\omega _{2})/2$,
temperature difference $\Delta T=T_{1}-T_{2}$, and energy detuning
$\Delta \omega=\omega _{1}-\omega _{2}$. And $\Delta \omega>0$ and
$\Delta \omega<0$ mean the positive and negative detunings,
respectively. For simplicity, in the following considerations we
assume $\gamma_{1}=\gamma_{1}=\gamma$.

In the following we consider three special cases: (1) The resonant
case, in which the two TLSs have the same energy separations, i.e.,
$\omega_{1}=\omega_{2}=\omega_{m}$, that is $\Delta\omega=0$. Now
the mixing angle $\theta=\pi/2$ and the energy separation
$\varepsilon=2\xi$. From Eq.~(\ref{probability2}) we obtain
\begin{eqnarray}
P_{\textrm{res}}\left( t\right)=\frac{1}{2}-\frac{1}{2}\cos(2\xi
t)e^{-\frac{1}{2}N(2\xi)\gamma t}, \label{resprobability}
\end{eqnarray}
where we introduce the parameter
\begin{eqnarray}
N(2\xi)=\bar{n}_{1}(2\xi)+\bar{n}_{2}(2\xi)+1.
\end{eqnarray}
The subscript ``\textrm{res}" stands for resonant case.
Equation~(\ref{resprobability}) means that the probability
$P_{\textrm{res}}$ increases from an initial value $0$ to a
steady-state value $1/2$ as the time $t$ increases. However, the
increase of the probability is exponential modulated by a cosine
function rather than monotone. In the short time limit it may
experience small oscillation.
\begin{figure}[tbp]
\includegraphics[width=8.6 cm]{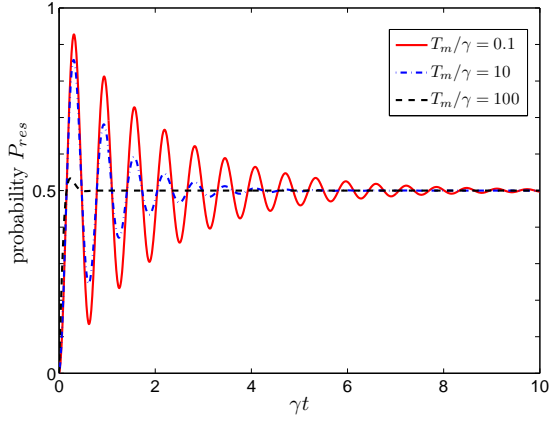}
\caption{(Color online) The probability $P_{\textrm{res}}$ given in
Eq.~(\ref{resonantprobability}) is plotted vs the scaled time
$\gamma t$ for different bath temperatures $T_{m}/\gamma=0.1$ (solid
red line), $10$ (dash dotted blue line), and $100$ (dashed black
line) in the resonant case $\Delta\omega/\gamma=0$. Other parameters
are set as $\gamma=1$, $\xi/\gamma=5$, and $\Delta
T/\gamma=0$.}\label{resonantprobability}
\end{figure}
The exponential rate $N(2\xi)\gamma/2$ is a function of the
parameters $\xi$, $\gamma$, $T_{1}$, and $T_{2}$. Obviously, the
parameter $N(2\xi)$ increases with the increase of the temperatures
of the heat baths. In the low temperature limit, i.e.,
$T_{1}/(2\xi)\approx0$ and $T_{2}/(2\xi)\approx0$, we have
$\bar{n}_{1}(2\xi)\approx0$ and $\bar{n}_{2}(2\xi)\approx0$, then
$N(2\xi)\approx1$. On the contrary, in the high temperature limit,
i.e., $T_{1}/(2\xi)\gg1$ and $T_{2}/(2\xi)\gg1$, we have
$\bar{n}_{1}(2\xi)\approx T_{1}/(2\xi)$ and
$\bar{n}_{2}(2\xi)\approx T_{2}/(2\xi)$, then
\begin{eqnarray}
N(2\xi)\approx\frac{T_{1}+T_{2}}{2\xi}+1\approx\frac{T_{m}}{\xi}.
\end{eqnarray}
The above equation means that in the high temperature limit, the
rate $N(2\xi)$ is proportional to the mean temperature $T_{m}$ and
does not depend on the temperature difference $\Delta T$. In
Fig.~\ref{resonantprobability}, we plot the probability
$P_{\textrm{res}}$ vs the scaled time $\gamma t$ for different bath
temperatures $T_{m}$, here we assume that $T_{1}=T_{2}=T_{m}$. From
Fig.~\ref{resonantprobability}, we can see that in the low
temperature limit the probability increases with an initial
oscillation. With the increase of the bath temperatures, the
oscillation disappears gradually.

\begin{figure}[tbp]
\includegraphics[width=8.6 cm]{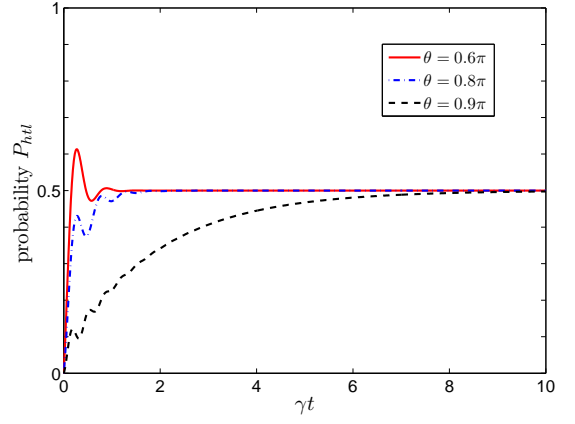}
\caption{(Color online) The probability $P_{\textrm{htl}}$ given in
Eq.~(\ref{diaprobforhighT}) vs the scaled time $\gamma t$ for
different mixing angle $\theta=0.6\pi$ (solid red line), $0.8\pi$
(dash dotted blue line), and $0.9\pi$ (dashed black line) at the
high temperature limit $T_{m}/\gamma=100$. Other parameters are set
as $\gamma=1$, $\xi/\gamma=5$,
$\chi_{1}/\gamma=\chi_{2}/\gamma=0.01T_{m}$, and $\Delta
T/\gamma=0$.}\label{highTprobability}
\end{figure}

(2) The high temperature limit case, i.e.,
$T_{1},T_{2}\gg\varepsilon$. In this case,
$\bar{n}_{1}(\varepsilon),\bar{n}_{2}(\varepsilon)\gg1$, then we can
make the approximations
$\bar{n}_{1}(\varepsilon)\approx\bar{n}_{1}(\varepsilon)+1$ and
$\bar{n}_{2}(\varepsilon)\approx\bar{n}_{2}(\varepsilon)+1$, which
lead to $\Gamma_{23}\approx\Gamma_{32}$. Therefore from
Eq.~(\ref{probability2}) we can obtain the time dependent
probability
\begin{eqnarray}
P_{\textrm{htl}}\left(
t\right)&\approx&\frac{1}{2}-\frac{1}{2}\cos^{2}\theta
e^{-\sin^{2}\theta N(\varepsilon)\gamma
t}\nonumber\\&&-\frac{1}{2}\sin ^{2}\theta \cos(\varepsilon
t)e^{-\left(2\cos^{2}\theta\chi+\frac{1}{2}\sin^{2}\theta
N(\varepsilon)\gamma\right)t},\label{diaprobforhighT}
\end{eqnarray}
where we introduce the parameter
$N(\varepsilon)=\bar{n}_{1}(\varepsilon)+\bar{n}_{2}(\varepsilon)+1$
and the subscript ``htl" stands for the high temperature limit.
Obviously, the above probability $P_{\textrm{htl}}$ increases from
an initial value $0$ to a steady-state value $1/2$. And the increase
of $P_{\textrm{htl}}$ is not simply exponential. In
Fig.~\ref{highTprobability}, we plot the probability
$P_{\textrm{htl}}$ vs the scaled time $\gamma t$ and the mixing
angle $\theta$ in the high temperature limit. Since the
probability~(\ref{diaprobforhighT}) is a function of
$\sin^{2}\theta$ and $\cos^{2}\theta$, therefore in
Fig.~\ref{highTprobability} we only need to plot the probability in
Eq.~(\ref{diaprobforhighT}) for the negative detuning cases.
Figure~\ref{highTprobability} shows that in the long time limit the
probability reaches $1/2$ irrespective of the $\theta$. Note that
here the mixing angle $0<\theta<\pi$. The cases of $0<\theta<\pi/2$
and $\pi/2<\theta<\pi$ mean the energy detuning $\Delta\omega>0$ and
$\Delta\omega<0$, respectively. And the angle $\theta=\pi/2$
corresponds to the resonant case. Here we choose
$0.1\pi<\theta<0.9\pi$, which corresponds to
$6.2>\Delta\omega/\xi>-6.2$.

\begin{figure}[tbp]
\includegraphics[width=8.2 cm]{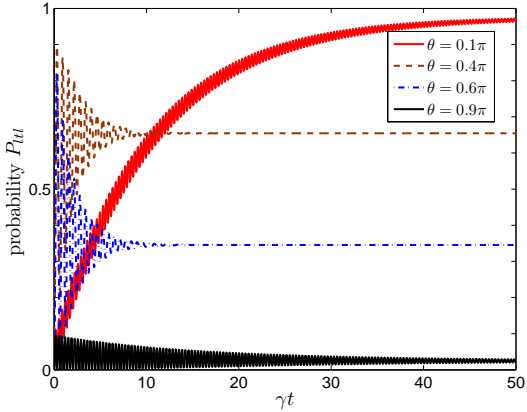}
\caption{(Color online) The probability $P_{\textrm{ltl}}(t)$ given
in Eq.~(\ref{lowtlimitprobability}) vs the scaled time $\gamma t$
for different mixing angle $\theta=0.1\pi$ (solid red line),
$0.4\pi$ (dashed brown line), $0.6\pi$ (dash dotted blue line), and
$0.9\pi$ (solid black line) at the low temperature limit
$T_{m}/\gamma=1$. Other parameters are set as $\gamma=1$,
$\xi/\gamma=5$, and $\Delta T/\gamma=0$.}\label{lowTprobability}
\end{figure}

(3) The low temperature limit case, i.e., $T_{1},T_{2}\approx0$. Now
we can make the approximations $\bar{n}_{1}(\varepsilon)\approx0$
and $\bar{n}_{2}(\varepsilon)\approx0$, which lead to
$\Gamma_{32}\approx0$ and
$\Gamma_{23}\approx\sin^{2}\theta~\gamma/2$. Then we obtain the
probability
\begin{eqnarray}
P_{\textrm{ltl}}(t)&\approx&\cos^{2}(\theta/2)\left(1-\cos\theta
e^{-\sin^{2}\theta\gamma
t}\right)\nonumber\\&&-\frac{1}{2}\sin^{2}\theta\cos(\varepsilon t)
e^{-\frac{1}{2}\sin^{2}\theta\gamma t},\label{lowtlimitprobability}
\end{eqnarray}
where the subscript ``\textrm{ltl}" means the low temperature limit.
In this case, the probability increases for an initial value $0$ to
a steady-state value $\cos^{2}(\theta/2)$. In
Fig.~\ref{lowTprobability}, we plot the probability
$P_{\textrm{ltl}}$ vs the time $t$ for different mixing angles
$\theta$ in the low temperature limit. Figure~\ref{lowTprobability}
shows that the probability $P_{\textrm{ltl}}$ increases from $0$ to
a steady state value with the increase of the time $t$. In the short
time, the probability experiences small oscillation. The steady
state value decreases with the increase of the $\theta$. Actually,
the obtained results are very reasonable from the viewpoint of
energy conservation. For the case of $\theta<\pi/2$, the energy
detuning $\Delta\omega>0$, we have $\omega_{1}>\omega_{2}$, then the
energy emitted by TLS$1$ can excite more than one TLS$2$ into their
excited state; For the case of $\theta>\pi/2$, we have
$\Delta\omega<0$, we have $\omega_{1}<\omega_{2}$, then the energy
emitted by TLS$1$ can only excite less than one TLS$2$ into the
excited state. Therefore, it is understandable that the steady-state
value of probability in low temperature increases as the parameter
$\theta$ decreases.

\subsection{Steady state case}

At steady state, the probability~(\ref{probability2}) becomes
\begin{eqnarray}
P_{ss}=\frac{1}{2}\left(1+\frac{\cos\theta}{N(\varepsilon)}\right),\label{steadystateprobabilityeq}
\end{eqnarray}
where the subscript ``ss" stands for steady state and
$N(\varepsilon)=\bar{n}_{1}(\varepsilon)+\bar{n}_{2}(\varepsilon)+1$.
This steady-state probability is a very interesting result since it
depends on the mixing angle $\theta$ and the bath temperatures
$T_{1}$ and $T_{2}$ independently. It depends on the mixing angle
$\theta$ and the bath temperatures $T_{1}$ and $T_{2}$ by
$\cos\theta$ and $1/N(\varepsilon)$, respectively.

\begin{figure}[tbp]
\includegraphics[width=8.6 cm]{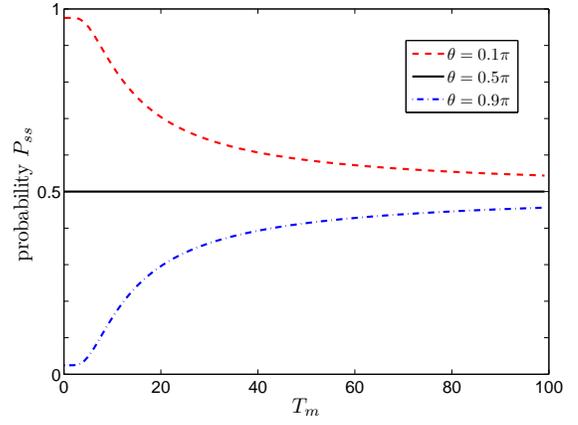}
\caption{(Color online) The steady state probability $P_{ss}$ vs the
bath temperature $T_{m}$ for different mixing angle $\theta=0.1\pi$
(dashed red line), $0.5$ (solid black line), and $0.9\pi$ (dash
dotted blue line). Other parameters are set as $\gamma=1$,
$\xi/\gamma=5$, and $\Delta
T/\gamma=0$.}\label{steadystateprobability-Tm}
\end{figure}

\begin{figure}[tbp]
\includegraphics[width=8.6 cm]{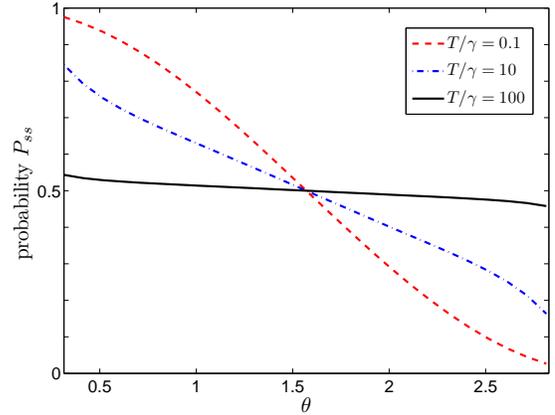}
\caption{(Color online) The steady state probability $P_{ss}$ vs the
mixing angle $\theta$ for different bath temperature
$T_{m}/\gamma=0.1$ (dashed red line), $10$ (dash dotted blue line),
and $100$ (solid black line). Other parameters are set as
$\gamma=1$, $\xi/\gamma=5$, and $\Delta
T/\gamma=0$.}\label{steadystateprobability-theta}
\end{figure}

We first consider several special cases at steady state: (1) The
resonant case, i.e., $\Delta\omega=0$. In this case, $\cos\theta=0$,
then $P_{ss}=1/2$. In the resonant case, the steady-state
probability $P_{ss}$ for single-excitation energy transfer is
independence of the temperatures of the two heat baths. This result
can also be understood from the following viewpoints: When
$\sin(\theta/2)=\cos(\theta/2)=1/\sqrt{2}$, the eigenstates
$|\lambda_{2}\rangle$ and $|\lambda_{3}\rangle$ become
$|\lambda_{2}\rangle=(|eg\rangle+|ge\rangle)/\sqrt{2}$ and
$|\lambda_{3}\rangle=(-|eg\rangle+|ge\rangle)/\sqrt{2}$. Therefore
for any statistical mixture
$\rho_{ss}=p_{2}\sigma_{22}+p_{3}\sigma_{33}$ of the two eigenstates
$|\lambda_{2}\rangle$ and $|\lambda_{3}\rangle$, the probability for
finding the two TLSs in state $|ge\rangle$ is $1/2$, where
$p_{2}+p_{3}=1$ is the normalization condition. (2) The high
temperature limit, i.e., $T_{1},T_{2}\gg\varepsilon$. In this case,
$\bar{n}_{1}(\varepsilon)\gg1$ and $\bar{n}_{2}(\varepsilon)\gg1$,
therefore $N(\varepsilon)\gg1$, which leads to $P_{ss}\approx1/2$.
In fact, in the high temperature limit, the steady state of the TLSs
should be $\rho_{s}\approx(\sigma_{22}+\sigma_{33})/2$, therefore
according to Eq.~(\ref{eigenstates}) we know that the probability
for finding the two TLSs in state $|ge\rangle$ is $1/2$. (3) The low
temperature limit, i.e., $T_{1},T_{2}\ll\varepsilon$. In this case,
$\bar{n}_{1}(\varepsilon)\approx0$ and
$\bar{n}_{2}(\varepsilon)\approx0$, then $N(\varepsilon)\approx1$,
which means $P_{ss}=\cos^{2}(\theta/2)$. In
Fig.~\ref{steadystateprobability-Tm}, we plot the steady-state
probability $P_{ss}$ in Eq.~(\ref{steadystateprobabilityeq}) vs the
bath temperatures $T_{m}$.
Figure~\ref{steadystateprobability-Tm} shows that, for the positive
detuning case, i.e., $0<\theta<\pi/2$, the steady state probability
$P_{ss}$ decreases from $1$ to $1/2$, but for the negative detuning
case, i.e., $\pi/2<\theta<\pi$, the $P_{ss}$ increases from $0$ to
$1/2$. For the resonant case, the $P_{ss}$ is $1/2$ irrespectively
of the bath temperature $T_{1}=T_{2}=T_{m}$. In
Fig.~\ref{steadystateprobability-theta}, we plot the steady-state
probability $P_{ss}$ in Eq.~(\ref{steadystateprobabilityeq}) vs the
mixing angle $\theta$.
Figure~\ref{steadystateprobability-theta} shows that, in the high
temperature case, the $P_{ss}$ becomes approximately a fixed value
$1/2$ irrespective of the $\theta$. But in the low temperature case,
the steady state probability $P_{ss}$ decreases with the increase of
$\theta$. These results are consistent with the above analysis.
Therefore, in the low temperature limit, we can improve the
steady-state probability $P_{ss}$ via increasing the $\Delta
\omega$.

In the above discussions of the steady-state probability, we have
assumed the bath temperature difference $\Delta T$ is zero.
Actually, we also study the dependence of the steady-state
probability on the bath temperature difference $\Delta T$ in both
the low and the high temperature limits. We found that the
dependence of the probability on $\Delta T$ is negligibly small with
the current parameters. This result is well understood from the
following viewpoint: in the low temperature limit, we have
$T_{1},T_{2}\ll\varepsilon$, therefore
$\bar{n}_{1}(\varepsilon)\approx0$ and
$\bar{n}_{2}(\varepsilon)\approx0$, $N(\varepsilon)\approx1$, then
$P_{ss}=\cos^{2}(\theta/2)$, which does not depends on the bath
temperature difference $\Delta T$; On the other hand, in the high
temperature limit, $T_{1},T_{2}\gg\varepsilon$, therefore
$\bar{n}_{1}(\varepsilon)\gg1$ and $\bar{n}_{2}(\varepsilon)\gg1$,
then
\begin{eqnarray}
P_{ss}\approx\frac{1}{2}\left(1+\frac{\varepsilon\cos\theta}{2T_{m}}\right),
\end{eqnarray}
which is independent of the bath temperature difference $\Delta T$.

\section{\label{Sec:5}Quantum entanglement between the donor and acceptor}

In this section, we study the quantum entanglement between the donor
and the acceptor with concurrence, which will be defined below. For
a $2\times2$ quantum system (two TLSs) with density matrix $\rho$
expressed in the bare state representation, its concurrence is
defined as~\cite{Wootters1998}
\begin{eqnarray}
C(\rho)=\max\{0,\sqrt{s_{1}}-\sqrt{s_{2}}-\sqrt{s_{3}}-\sqrt{s_{4}}\},
\end{eqnarray}
where $s_{i}$ ($i=1,2,3,4$) are the eigenvalues ($s_{1}$ being the
largest one) of the matrix $\rho\tilde{\rho}$, where the operator
$\tilde{\rho}$ is define as
\begin{eqnarray}
\tilde{\rho}=(\sigma^{y}_{1}\otimes\sigma^{y}_{2})\rho^{\ast}(\sigma^{y}_{1}\otimes\sigma^{y}_{2})
\end{eqnarray}
with $\rho^{\ast}$ being the complex conjugate of $\rho$. Note that
here $\sigma^{y}_{i}$ is the usual Pauli matrix pointing the $y$
axis. For the $2\times2$ quantum system, the concurrences $C=0$ and
$C=1$ mean the density matrix $\rho$ is an unentangled and maximally
entangled states, respectively. Specially, for the ``X"-class state
with the density matrix
\begin{eqnarray}
\rho=\left(
       \begin{array}{cccc}
         \rho_{11} &0 & 0 & \rho_{14} \\
         0 & \rho_{22} & \rho_{23} & 0 \\
         0 & \rho_{32} & \rho_{33} & 0 \\
         \rho_{41} & 0 & 0 & \rho_{44} \\
       \end{array}
     \right)
\end{eqnarray}
expressed in the bare state representation, the concurrence
is~\cite{Zubairy1998}
\begin{eqnarray}
C(\rho)=\max\{0,2(|\rho_{23}|-\sqrt{\rho_{11}\rho_{44}}),2(|\rho_{14}|-\sqrt{\rho_{22}\rho_{33}})\}.\label{Xstateconcu}
\end{eqnarray}

Now, for the present system, its density matrix $\rho$ can be
expressed as the following form in the bare state representation,
\begin{eqnarray}
\rho=\left(
       \begin{array}{cccc}
         \langle\tau_{11}\rangle & \langle\tau_{21}\rangle & \langle\tau_{31}\rangle & \langle\tau_{41}\rangle \\
         \langle\tau_{12}\rangle & \langle\tau_{22}\rangle & \langle\tau_{32}\rangle & \langle\tau_{42}\rangle \\
         \langle\tau_{13}\rangle & \langle\tau_{23}\rangle & \langle\tau_{33}\rangle & \langle\tau_{43}\rangle \\
         \langle\tau_{14}\rangle & \langle\tau_{24}\rangle & \langle\tau_{34}\rangle & \langle\tau_{44}\rangle \\
       \end{array}
     \right),
\end{eqnarray}
where the density matrix elements are defined as
$\langle\tau_{ij}\rangle=\textrm{Tr}[\tau_{ij}\rho]=\textrm{Tr}[|\eta_{i}\rangle\langle\eta_{j}|\rho]
=\langle\eta_{j}|\rho|\eta_{i}\rangle$
with the transition operator
$\tau_{ij}=|\eta_{i}\rangle\langle\eta_{j}|$. Since the concurrence
is defined in the bare state representation and the evolution of the
system is expressed in the eigenstate representation. Therefore we
need to obtain the transformation between the two representations.
The density matrix elements in the eigenstate and bare state
representations are expressed by $\langle\sigma _{ij}(t)\rangle$ and
$\langle \tau _{ij}(t)\rangle$, respectively. Making using of
Eq.~(\ref{eigenstates}), we can obtain the relations for diagonal
density matrix elements
\begin{eqnarray}
\left\langle \sigma _{11}(t)\right\rangle &=&\left\langle \tau
_{11}(t)\right\rangle,\hspace{0.5 cm} \left\langle \sigma
_{44}(t)\right\rangle =\left\langle \tau
_{44}(t)\right\rangle,\nonumber\\
\left\langle \sigma _{22}(t)\right\rangle &=&\cos ^{2}\left( \theta
/2\right) \left\langle \tau _{22}(t)\right\rangle +\sin ^{2}\left(
\theta /2\right) \left\langle \tau _{33}(t)\right\rangle\nonumber\\
&&+\frac{1}{2}\sin \theta \left( \left\langle \tau
_{23}(t)\right\rangle +\left\langle \tau_{32}(t)\right\rangle
\right),\nonumber\\
\left\langle \sigma _{33}(t)\right\rangle &=&\sin ^{2}\left( \theta
/2\right) \left\langle \tau _{22}(t)\right\rangle +\cos ^{2}\left(
\theta /2\right) \left\langle \tau _{33}(t)\right\rangle\nonumber\\
&&-\frac{1}{2}\sin \theta \left( \left\langle \tau
_{23}(t)\right\rangle +\left\langle \tau _{32}(t)\right\rangle
\right),\label{tansformation}
\end{eqnarray}
and the following off-diagonal element which will be useful below,
\begin{eqnarray}
 \left\langle \sigma _{23}(t)\right\rangle
&=&\frac{1}{2}\sin \theta (\left\langle \tau
_{33}(t)\right\rangle-\left\langle \tau
_{22}(t)\right\rangle)\nonumber\\
&&+\cos ^{2}\left( \theta /2\right) \left\langle \tau
_{23}(t)\right\rangle -\sin ^{2}\left( \theta /2\right) \left\langle
\tau _{32}(t)\right\rangle.\label{reprransformation}
\end{eqnarray}
Correspondingly, we can obtain the inverse transform
\begin{eqnarray}
\left\langle \tau _{22}(t)\right\rangle &=&\cos ^{2}\left( \theta
/2\right) \left\langle \sigma _{22}(t)\right\rangle +\sin ^{2}\left(
\theta /2\right) \left\langle \sigma
_{33}(t)\right\rangle\nonumber\\
&&-\frac{1}{2}\sin \theta \left( \left\langle \sigma
_{23}(t)\right\rangle +\left\langle \sigma
_{32}(t)\right\rangle \right),\nonumber\\
\left\langle \tau _{33}(t)\right\rangle&=&\sin ^{2}\left( \theta
/2\right) \left\langle \sigma _{22}(t)\right\rangle +\cos ^{2}\left(
\theta /2\right) \left\langle \sigma _{33}(t)\right\rangle\nonumber\\
&&+\frac{1}{2}\sin \theta \left( \left\langle \sigma
_{23}(t)\right\rangle +\left\langle \sigma
_{32}(t)\right\rangle \right),\nonumber\\
\left\langle \tau _{23}(t)\right\rangle &=&-\sin ^{2}\left( \theta
/2\right) \left\langle \sigma _{32}(t)\right\rangle +\cos ^{2}\left(
\theta /2\right) \left\langle \sigma _{23}(t)\right\rangle\nonumber\\
&&+\frac{1}{2}\sin \theta \left( \left\langle \sigma
_{22}(t)\right\rangle -\left\langle \sigma _{33}(t)\right\rangle
\right).\label{reprransf}
\end{eqnarray}
Also here we only express explicitly the elements which will be used
below.

In order to calculate the concurrence of the system, we need to know
its density matrix in the bare representation for a given initial
state. Fortunately, the evolution relation from $\langle\tau
_{ij}(0)\rangle$ to $\langle\tau _{ij}(t)\rangle$ can be obtained
through the following process
\begin{eqnarray}
\langle\tau _{ij}(0)\rangle\rightarrow\langle \sigma _{ij}(0)\rangle
\rightarrow\langle \sigma _{ij}(t)\rangle\rightarrow\langle \tau
_{ij}(t)\rangle.\label{relation}
\end{eqnarray}
Concretely, the transformation relations $\langle\tau
_{ij}(0)\rangle\rightarrow\langle \sigma _{ij}(0)\rangle$ and
$\langle \sigma _{ij}(t)\rangle\rightarrow\langle \tau
_{ij}(t)\rangle$ are determined by Eqs.~(\ref{tansformation}), (\ref{reprransformation})
and~(\ref{reprransf}), and the evolution relation $\langle \sigma
_{ij}(0)\rangle \rightarrow\langle \sigma _{ij}(t)\rangle$ is
determined by Eq.~(\ref{transientsolution2}). In terms of
Eqs.~(\ref{transientsolution2}),~(\ref{tansformation}),~(\ref{reprransformation}),
~(\ref{reprransf}), and~(\ref{relation}), we can obtain the
following relation
\begin{widetext}
\begin{eqnarray}
\left\langle \tau _{23}\left(
t\right) \right\rangle &=&\left[\frac{1}{2}\sin \theta \frac{\Gamma
_{32}-\Gamma _{23}}{\Gamma _{23}+\Gamma _{32}}+\sin \theta \frac{(
\cos ^{2}\left( \theta /2\right) \Gamma _{23}-\sin ^{2}\left( \theta
/2\right) \Gamma _{32})
}{\Gamma_{23}+\Gamma _{32}}e^{-2\left( \Gamma _{23}+\Gamma _{32}\right) t}\right.\nonumber\\
&&\left.-\frac{1}{2}\sin\theta e^{-\left( \cos ^{2}\theta \Pi
_{1}+\Gamma _{23}+\Gamma _{32}\right) t}\left( e^{i\varepsilon
t}\cos ^{2}\left( \theta /2\right)-e^{-i\varepsilon t}\sin
^{2}\left( \theta /2\right) \right)\right]\left\langle \tau
_{22}\left( 0\right)
\right\rangle\nonumber\\
&&+\left[\frac{1}{2}\sin \theta \frac{\Gamma _{32}-\Gamma
_{23}}{\Gamma _{23}+\Gamma _{32}}+\sin \theta \frac{\sin ^{2}\left(
\theta /2\right) \Gamma _{23}-\cos ^{2}\left( \theta /2\right)
\Gamma _{32}) }{\Gamma _{23}+\Gamma _{32}}e^{-2\left( \Gamma
_{23}+\Gamma
_{32}\right) t}\right.\nonumber\\
&&\left.+ \frac{1}{2}\sin(\theta) e^{-\left( \cos ^{2}\theta \Pi
_{1}+\Gamma _{23}+\Gamma _{32}\right) t}\left( e^{i\varepsilon
t}\cos ^{2}\left( \theta /2\right) -e^{-i\varepsilon t}\sin
^{2}\left( \theta /2\right) \right)\right]\left\langle \tau
_{33}\left(
0\right) \right\rangle\nonumber\\
&&+\left[(\sin ^{4}\left( \theta /2\right)e^{-i\varepsilon t}+\cos
^{4}\left( \theta /2\right)e^{i\varepsilon t}) e^{-\left( \cos
^{2}\theta\Pi_{1}+\Gamma _{23}+\Gamma _{32}\right)
t}+\frac{1}{2}\sin ^{2}\theta e^{-2\left( \Gamma _{23}+\Gamma
_{32}\right) t}\right]\left\langle \tau _{23}\left( 0\right) \right\rangle\nonumber\\
&&+\frac{1}{2}\sin ^{2}\theta \left( e^{-2\left( \Gamma _{23}+\Gamma
_{32}\right) t}-e^{-\left( \cos ^{2}\theta \Pi _{1}+\Gamma
_{23}+\Gamma _{32}\right) t}\cos(\varepsilon t)\right)\left\langle
\tau _{32}\left( 0\right) \right\rangle.\label{tmap}
\end{eqnarray}
\end{widetext}

Now, we obtain the evolution relation of the density matrix elements
in the bare state representation. Since the expressions are very
complex, here we only show the matrix elements which will be used in
the following. Based on these evolutionary matrix elements, we can
write out the density matrix of the system in bare state
representation at time $t$ once the initial state is given, and then
we can obtain the concurrence of the density matrix. In what
follows, we will discuss the entanglement dynamics and steady-state
entanglement.

\subsection{Entanglement dynamics}

In the process of single-excitation energy transfer from the donor
to the acceptor, the single excitation energy is initially possessed
by the donor and the acceptor is in its ground state. Therefore the
initial state of the system is
\begin{eqnarray}
|\psi(0)\rangle=|eg\rangle=|\eta_{2}\rangle,
\end{eqnarray}
which means the initial conditions are that all matrix elements are
zero except $\langle\tau_{22}(0)\rangle=1$. According to
Eq.~(\ref{tmap}), we know that the density matrix $\rho(t)$ of the
system belongs to the so-called $X$-class state. Then the
concurrence can be obtained with Eq.~(\ref{Xstateconcu})
\begin{eqnarray}
\label{transientconcurrence} C(t)&=&2\left|\left[\frac{1}{2}\sin
\theta \frac{\Gamma _{32}-\Gamma _{23}}{\Gamma _{23}+\Gamma
_{32}}\right.\right.\nonumber\\
&&\left.\left.+\sin \theta \frac{( \cos ^{2}\left( \theta /2\right)
\Gamma _{23}-\sin ^{2}\left( \theta /2\right) \Gamma _{32})
}{\Gamma_{23}+\Gamma _{32}}e^{-2\left( \Gamma _{23}+\Gamma _{32}\right) t}\right.\right.\nonumber\\
&&\left.\left.-\frac{1}{2}\sin\theta e^{-\left( \cos ^{2}\theta \Pi
_{1}+\Gamma _{23}+\Gamma _{32}\right) t}\right.\right.\nonumber\\
&&\left.\left.\times\left( e^{i\varepsilon t}\cos ^{2}\left( \theta
/2\right)-e^{-i\varepsilon t}\sin ^{2}\left( \theta /2\right)
\right)\right]\right|.
\end{eqnarray}

In what follows, we consider three special cases of interest: (1)
The resonant case, i.e., $\omega_{1}=\omega_{2}=\omega_{m}$, that is
$\Delta\omega=0$. Then the mixing angle $\theta=\pi/2$ and the
energy separation $\varepsilon=2\xi$, thus we obtain
\begin{eqnarray}
C_{\textrm{res}}(t)&\approx&\left|\frac{1}
{N(2\xi)}\left(1-e^{-N(2\xi)\gamma t}\right)+i\sin(\varepsilon
t)e^{-\frac{1}{2}N(2\xi)\gamma
t}\right|,\label{resonantconcurrenceeq}
\end{eqnarray}
where $N(2\xi)=\bar{n}_{1}(2\xi)+\bar{n}_{2}(2\xi)+1$. From
Eq.~(\ref{resonantconcurrenceeq}), we find that the concurrence
$C_{\textrm{res}}(t)$ increases from zero to a steady state value
$1/N(2\xi)$ with the increase of the time $t$. Clearly, the steady
state concurrence $1/N(2\xi)$ decreases from one to zero as the
temperature $T_{m}$ increases from zero to infinite.
\begin{figure}[tbp]
\includegraphics[width=8.6 cm]{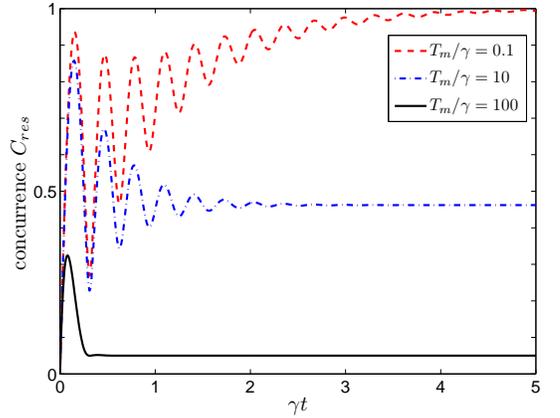}
\caption{(Color online) The concurrence $C_{\textrm{res}}$ in
Eq.~(\ref{resonantconcurrenceeq}) vs the scaled time $\gamma t$ for
different bath temperature $T_{m}/\gamma=0.1$ (dashed red line),
$10$ (dash dotted blue line), and $100$ (solid black line) in the
resonant case $\Delta\omega/\gamma=0$. Other parameters are set as
$\gamma=1$, $\xi/\gamma=5$, and $\Delta
T/\gamma=0$.}\label{resonantentanglement}
\end{figure}
In Fig.~\ref{resonantentanglement}, we plot the
concurrence~(\ref{resonantconcurrenceeq}) in the resonant case vs
the scaled time $\gamma t$ and for different heat bath average
temperatures $T_{m}$. Figure~\ref{resonantentanglement} shows the
results as we analyze above.

\begin{figure}[tbp]
\includegraphics[width=8.6 cm]{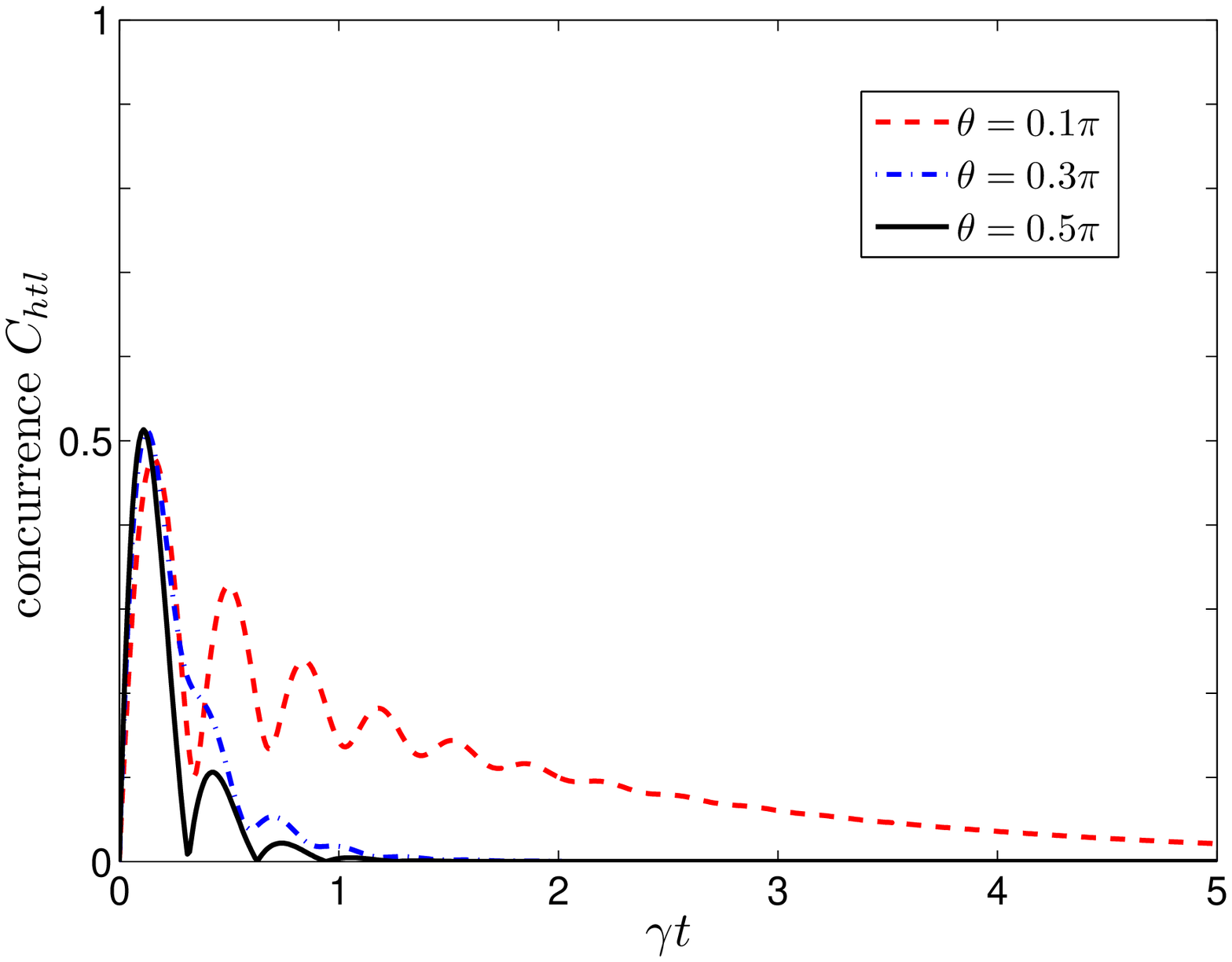}
\caption{(Color online) The concurrence $C_{\textrm{htl}}$ in
Eq.~(\ref{highTconcurrenceeq}) vs the scaled evolution time $\gamma
t$ for different mixing angle $\theta=0.1\pi$ (dashed red line),
$0.3\pi$ (dashed blue line), and $0.5\pi$ (solid black line) in the
high temperature limit $T_{m}/\gamma=100$. Other parameters are set
as $\gamma=1$, $\xi/\gamma=5$,
$\chi_{1}/\gamma=\chi_{2}/\gamma=0.01T_{m}$, and $\Delta
T/\gamma=0$.}\label{highTentanglement}
\end{figure}

(2) The high temperature limit, i.e., $T_{1},T_{2}\gg\varepsilon$.
In this case,
$\bar{n}_{1}(\varepsilon),\bar{n}_{2}(\varepsilon)\gg1$, then we can
have the approximate relations
$\bar{n}_{1}(\varepsilon)\approx\bar{n}_{1}(\varepsilon)+1$ and
$\bar{n}_{2}(\varepsilon)\approx\bar{n}_{2}(\varepsilon)+1$, which
lead to $\Gamma_{23}\approx\Gamma_{32}$. Then the
concurrence~(\ref{transientconcurrence}) becomes
\begin{eqnarray}
\label{highTconcurrenceeq}
C_{\textrm{htl}}(t)&\approx&\left|\frac{\sin(2\theta)}{2}e^{-\sin^{2}\theta
N(\varepsilon)\gamma t}-\sin\theta
e^{-\left(2\cos^{2}\theta\chi+\frac{1}{2}\sin^{2}\theta
N(\varepsilon)\gamma\right) t}\right.\nonumber\\
&&\left.\times\left( e^{i\varepsilon t}\cos ^{2}\left( \theta
/2\right)-e^{-i\varepsilon t}\sin ^{2}\left( \theta /2\right)
\right)\right|.
\end{eqnarray}
The expression of the concurrence~(\ref{highTconcurrenceeq}) in the
high temperature limit is not simple as that of the resonant case,
but we can still observe the two points: The first is that the
dependence of the concurrence on the angle $\theta$ is approximately
$\sin\theta$; and the second is that the steady-state concurrence is
zero, which means there is no quantum entanglement between the donor
and the acceptor. This result can also be seen from the density
operator of the steady state for the donor and the acceptor. In the
high temperature limit, the steady state density matrix of the donor
and the acceptor is $\rho\approx(|eg\rangle\langle
eg|+|ge\rangle\langle ge|)/2$, which is an unentangled state.
Physically, this result is direct since the quantum systems will
transit to classical systems in the high temperature limit. In
Fig.~\ref{highTentanglement}, we plot the concurrence given by
Eq.~(\ref{highTconcurrenceeq}) vs the evolution time $t$ for
different mixing angles $\theta$. Figure~\ref{highTentanglement}
shows that the concurrence experiences an increase from zero to a
maximal value and then decreases to a steady state value with the
scaled time $\gamma t$.

\begin{figure}[tbp]
\includegraphics[width=8.6 cm]{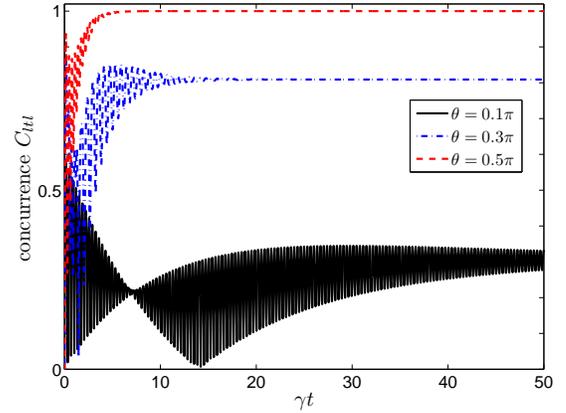}
\caption{(Color online) The concurrence $C_{\textrm{ltl}}$ in
Eq.~(\ref{lowTconcurrenceeq}) vs the scaled evolution time $\gamma
t$ for different mixing angle $\theta=0.1\pi$ (solid black line),
$0.3\pi$ (dash dotted blue line), and $0.5\pi$ (dashed red line) is
plotted in the low temperature limit $T_{m}/\gamma=0.01$. Other
parameters are set as $\gamma=1$, $\xi/\gamma=5$,
$\chi_{1}/\gamma=\chi_{2}/\gamma=0.01T_{m}$, and $\Delta
T/\gamma=0$.}\label{lowTentanglement}
\end{figure}

(3) The low temperature limit, i.e., $T\approx0$. Now we can
approximately have $\bar{n}(\varepsilon)\approx0$, which lead to
$\Gamma_{32}\approx0$ and
$\Gamma_{23}\approx\sin^{2}\theta\gamma/2$. Then the
concurrence~(\ref{transientconcurrence}) becomes
\begin{eqnarray}
C_{\textrm{ltl}}(t)&\approx&\sin
\theta\left|1-2\cos^{2}(\theta/2)e^{-\sin^{2}\theta\gamma
t}+e^{-\frac{1}{2}\sin^{2}\theta\gamma
t}\right.\nonumber\\&&\left.\times\left(e^{i\varepsilon t}\cos
^{2}\left( \theta /2\right)-e^{-i\varepsilon t}\sin ^{2}\left(
\theta /2\right) \right)\right|,\label{lowTconcurrenceeq}
\end{eqnarray}
where the subscript ``\textrm{ltl}" stands for low temperature
limit. Similar to the high temperature limit, the increase of the
concurrence is also not simply exponential. The concurrence
increases from zero to a steady state value $\sin\theta$ with the
increase of the scaled time $t$, which means the concurrence at long
time limit  is irrespective of the sign of the detuning. This long
lived entanglement is much larger than that of the high temperature
limit. We can also see the steady state concurrence from the
viewpoint of quantum noise. When $T\approx0$, the steady state of
the donor and the acceptor is
$\rho\approx|\lambda_{3}\rangle\langle\lambda_{3}|$ with concurrence
$\sin\theta$. In Fig.~\ref{lowTentanglement}, we plot the
concurrence given by Eq.~(\ref{lowTconcurrenceeq}) vs the evolution
time $t$ and the mixing angle $\theta$.
Figure~\ref{lowTentanglement} shows that the concurrence increases
from zero to a steady state value with the scaled time $t$.

\subsection{Steady state entanglement}

\begin{figure}[tbp]
\includegraphics[width=8.6 cm]{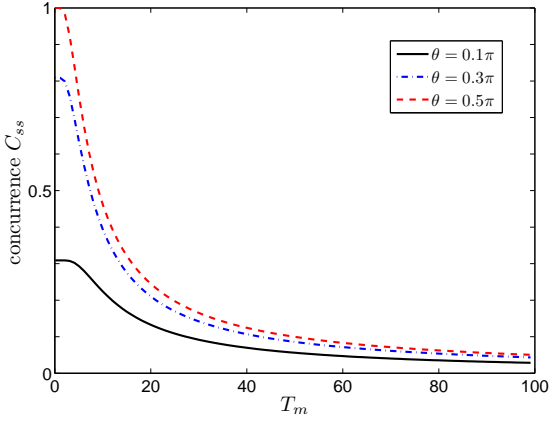}
\caption{(Color online) The steady state concurrence $C_{ss}$ vs the
bath temperature $T_{m}$ for different mixing angle $\theta=0.1\pi$
(solid black line), $0.3$ (dash dotted blue line), and $0.5\pi$
(dashed red line). Other parameters are set as $\gamma=1$,
$\xi/\gamma=5$, and $\Delta
T/\gamma=0$.}\label{steadystateentanglement-Tm}
\end{figure}

\begin{figure}[tbp]
\includegraphics[width=8.6 cm]{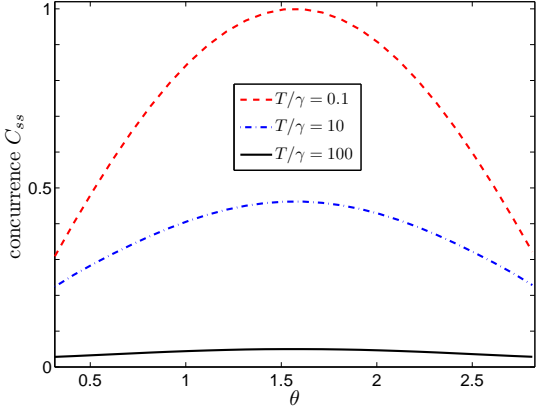}
\caption{(Color online) The steady state concurrence $C_{ss}$ vs the
mixing angle $\theta$ for different bath temperature
$T_{m}/\gamma=0.1$ (dashed red line), $10$ (dash dotted blue line),
and $100$ (solid black line). Other parameters are set as
$\gamma=1$, $\xi/\gamma=5$, and $\Delta
T/\gamma=0$.}\label{steadystateentanglement-theta}
\end{figure}

From Eq.~(\ref{transientconcurrence}), it is straightforward to
obtain the steady state concurrence between the donor and the
acceptor,
\begin{eqnarray}
\label{steadystateconcurrenceeq}
C_{ss}&=&\frac{\sin\theta}{N(\varepsilon)}.
\end{eqnarray}
In the high temperature limit, we have
$C_{\textrm{htl}}(\infty)\approx0$, and in the low temperature
limit, we have $C_{\textrm{ltl}}(\infty)\approx\sin \theta$. For a
general state, it is interesting to point out that the steady state
concurrence $C_{ss}$ depends on the temperature $T_{m}$ and the
angle $\theta$ independently. For a given $\theta$, the dependence
on $T_{m}$ is inverse proportional to $N(\varepsilon)$, and for a
given $T_{m}$, the dependence on $\theta$ is $\sin\theta$. In
Fig.~\ref{steadystateentanglement-Tm}, we plot the concurrence given
by Eq.~(\ref{steadystateconcurrenceeq}) vs the temperature $T_{m}$
for different mixing angles $\theta$.
Figure~\ref{steadystateentanglement-Tm} shows that the steady state
concurrence decreases with the increase of the temperature $T_{m}$.
In Fig.~\ref{steadystateentanglement-theta}, we plot the concurrence
given by Eq.~(\ref{steadystateconcurrenceeq}) vs the mixing angle
$\theta$ for different average bath temperature $T_{m}$.
Figure~\ref{steadystateentanglement-theta} shows that the dependence
of the concurrence on the mixing angle $\theta$ decreases with the
increase of the average bath temperature $T_{m}$. Moreover, from
Eq.~(\ref{steadystateconcurrenceeq}), we can also see that the
steady-state concurrence is independent of $\Delta T$ at the high
temperature limit.

\section{\label{Sec:6}concluding with remarks}

In conclusion, we have studied analytically coherent
single-excitation energy transfer in a dimer consisting of a donor
and an acceptor modeled by two TLSs, which are immersed in two
independent heat baths. Special attention is paid to the effect on
the single-excitation energy transfer probability of the energy
detuning and the heat bath temperatures of the two TLSs. It has been
found that, the probability for single-excitation energy transfer
largely depends on the energy detuning in the low temperature limit.
Concretely, the positive and negative energy detunings can increase
and decrease the probability, respectively. In the high temperature
limit, however, the effect of the energy detuning on the probability
is negligibly small. We have also found that the probability is
negligibly dependence on the bath temperature difference in the low
and high temperature limits. We have also studied analytically
quantum entanglement in the dimer system through calculating quantum
concurrence. It was found that quantum entanglement can be created
during the process of excitation energy transfer. The steady state
entanglement between the donor and the acceptor decreases with the
increasing of the bath temperature. And the dependence of the steady
state concurrence on the energy detuning is proportional to the sine
function of the mixing angle and irrespective of the bath
temperatures. Moreover, we have found that the dependence of the
steady state concurrence on the bath temperature difference is
negligibly small with the current parameters.

Finally, we give two remarks on the above obtained results: First,
we should distinguish the present work from dynamic disentanglement
suddenly or asymptotically (e.g.,
Refs.~\cite{YuTing,Almeida,Ficek,FQWang,Dubi,BAn,Ann,ZheSun,
HZheng,James,Bellomo,Davidovich,Ban}). Mainly, there are three
points of difference between the two cases: the initial state, the
coupling between the two TLSs, and the coupling form between the
TLSs and their heat baths. In dynamic disentanglement, the two TLSs
is initially prepared in an entanglement state, there is no coupling
between the two TLSs, and the coupling form of the TLSs with their
heat baths is off-diagonal. But in the present work, initially the
two TLSs are unentangled, there is a dipole-dipole interaction
between the two TLSs, and the coupling form of the TLSs with their
heat baths is diagonal. Certainly, the results also differ. In
entanglement sudden death, the two TLSs disentangle to zero
suddenly. But in this work, steady state entanglement is created.

Second, in this work, we only address the problem about how is the
dynamics of the \textit{created} quantum entanglement in the process
of excitation energy transfer~\cite{Scholak}. But we do not address
the question about the relation between \textit{initially prepared}
quantum entanglement among the pigments and the efficiency for
single-excitation energy transfer. Just as in quantum information
science, quantum entanglement is considered an important resource
since it can be used to enhance the efficiency of quantum
information protocols. Therefore it remains a question whether
initially prepared quantum entanglement can enhance the efficiency
of excitation energy transfer.

\acknowledgments
This work is supported in part by NSFC Grants
No.~10935010 and No.~10775048, NFRPC Grants No.~2006CB921205 and
No.~2007CB925204.

\appendix*
\section{\label{appenfordiagonal}Derivation of quantum master
equation~(\ref{mastereqfordiagonalcase})}

In this appendix, we present a detailed derivation of quantum master
equation~(\ref{mastereqfordiagonalcase}). Let us start from the
Hamiltonian~(\ref{Hamiltonian}) of the total system. In the
interaction picture with respect to the Hamiltonian
\begin{eqnarray}
H_{0}=H_{\textrm{TLSs}}+H_{B},
\end{eqnarray}
the interaction
Hamiltonian~(\ref{diacouplingH}) becomes
\begin{eqnarray}
H_{I}(t)&=&\sigma_{11}B_{11}(t)+\sigma_{22}B_{22}(t)+\sigma_{33}B_{33}(t)\nonumber\\
&&+\sigma_{23}e^{i\varepsilon
t}B_{23}(t)+\sigma_{32}e^{-i\varepsilon
t}B^{\dag}_{23}(t),\label{HIininteractionpict}
\end{eqnarray}
through introducing the following noise operators:
\begin{eqnarray}
\label{noiseoperators}
B_{11}(t)&=&A(t)+B(t),\nonumber\\
B_{22}(t)&=&\cos^{2}(\theta/2)A(t)+\sin^{2}(\theta/2)B(t),\nonumber\\
B_{33}(t)&=&\sin^{2}(\theta/2)A(t)+\cos^{2}(\theta/2)B(t),\nonumber\\
B_{23}(t)&=&\frac{\sin\theta}{2}\left(\sum_{k}g_{2k}b_{k}e^{-i\omega
_{bk}t}-\sum_{j}g_{1j}a_{j}e^{-i\omega _{aj}t}\right),
\end{eqnarray}
with
\begin{eqnarray}
A(t)&\equiv&e^{iH^{(a)}_{B}t}A(0)e^{-iH^{(a)}_{B}t}\nonumber\\
&=&\sum_{j}g_{1j}(a_{j}^{\dagger
}e^{i\omega
_{aj}t}+a_{j}e^{-i\omega _{aj}t}),\nonumber\\
B(t)&\equiv&e^{iH^{(b)}_{B}t}B(0)e^{-iH^{(b)}_{B}t}\nonumber\\
&=&\sum_{k}g_{2k}(b_{k}^{\dagger}e^{i\omega
_{bk}t}+b_{k}e^{-i\omega _{bk}t}).\label{aboperators}
\end{eqnarray}
Obviously, $A(t)$ and $B(t)$ are Hermitian operators. Note that in
Eq.~(\ref{HIininteractionpict}) we have made rotating wave
approximation.

Under the Born-Markov approximation, the master equation
reads~\cite{Breuer}
\begin{eqnarray}
\dot{\rho}_{S}=-\int_{0}^{\infty}d\tau\textrm{Tr}_{B}[H_{I}(t),[H_{I}(t-\tau),\rho_{S}\otimes\rho_{B}]],\label{mastereqformula}
\end{eqnarray}
where $\textrm{Tr}_{B}$ stands for tracing over the degrees of
freedom of the heat baths. The density matrix
$\rho_{B}\equiv\rho^{(a)}_{th}\otimes\rho^{(b)}_{th}$ of the heat
baths means the two independent heat baths being in thermal
equilibrium,
\begin{eqnarray}
\rho^{(a)}_{th}&=&Z^{-1}_{A}\exp(-\beta_{1}H^{(a)}_{B}),\nonumber\\
\rho^{(b)}_{th}&=&Z^{-1}_{B}\exp(-\beta_{2}H^{(b)}_{B}),\label{thermstdensitymatrix}
\end{eqnarray}
where we denote the partition functions
$Z_{A}=\textrm{Tr}_{B_{a}}[\exp(-\beta_{1}H^{(a)}_{B})]$ and
$Z_{B}=\textrm{Tr}_{B_{b}}[\exp(-\beta_{2}H^{(b)}_{B})]$ with
$\beta_{1}=1/T_{1}$ and $\beta_{2}=1/T_{2}$ being respectively the
inverse temperatures of the heat baths of the TLS$1$ and TLS$2$.

By using Eqs.~(\ref{HIininteractionpict})
and~(\ref{mastereqformula}) and making rotating wave approximation,
we can obtain the following quantum master equation
\begin{eqnarray}
\label{mastereq}
\dot{\rho}_{S}&=&\sigma_{32}\rho_{S}\sigma_{23}\int_{0}^{\infty}e^{-i\varepsilon
t}\langle B_{23}(-\tau)B^{\dag}_{23}(0)\rangle
d\tau\nonumber\\
&&+\sigma_{23}\rho_{S}\sigma_{32}\int_{0}^{\infty}e^{i\varepsilon
t}\langle B^{\dag}_{23}(-\tau)B_{23}(0)\rangle d\tau\nonumber\\
&&-\sigma_{22}\rho_{S}\int_{0}^{\infty}e^{i\varepsilon t}\langle
B_{23}(0)B^{\dag}_{23}(-\tau)\rangle
d\tau\nonumber\\
&&-\sigma_{33}\rho_{S}\int_{0}^{\infty}e^{-i\varepsilon t}\langle
B^{\dag}_{23}(0)B_{23}(-\tau)\rangle
d\tau\nonumber\\
&&+\sum_{m,n=1,2,3}\sigma_{mm}\rho_{S}\sigma_{nn}\int_{0}^{\infty}\langle
B_{nn}(-\tau)B_{mm}(0)\rangle
d\tau\nonumber\\
&&-\sum_{n=1,2,3}\sigma_{nn}\rho_{S}\int_{0}^{\infty}\langle
B_{nn}(0)B_{nn}(-\tau)\rangle d\tau\nonumber\\
&&+h.c.,
\end{eqnarray}
where the correlation functions for the bath operators are defined
as $\langle
X(t)Y(t')\rangle\equiv\textrm{Tr}_{B}[X(t)Y(t')\rho_{B}]$. Notice
that here we use the property $\langle X(t)Y(t')\rangle=\langle
X(t-t')Y(0)\rangle=\langle X(0)Y(t'-t)\rangle$ of the correlation
functions for the bath operators. To derive the quantum master
equation we need to calculate the Fourier transform of the
correlation functions in Eq.~(\ref{mastereq}). For simplicity, here
we only keep the real parts of the Fourier transforms of the
correlation functions and neglect their imaginary parts since the
imaginary parts only contribute to the Lamb shifts, which are
neglected in this work. According to
Eqs.~(\ref{noiseoperators}),~(\ref{aboperators}),
and~(\ref{thermstdensitymatrix}), we can express the Fourier
transforms for the correlation functions in Eq.~(\ref{mastereq}) as
follows:
\begin{eqnarray}
\int_{0}^{\infty }d\tau\langle
B_{11}(0)B_{11}(-\tau)\rangle&=&\int_{0}^{\infty }d\tau\langle
B_{11}(-\tau)B_{11}(0)\rangle^{\ast}\nonumber\\
&=&D_{A}
+D_{B},\nonumber\\
\int_{0}^{\infty }d\tau\langle
B_{22}(0)B_{22}(-\tau)\rangle&=&\int_{0}^{\infty }d\tau\langle
B_{22}(-\tau)B_{22}(0)\rangle^{\ast}\nonumber\\
&=&\cos^{4}(\theta/2)D_{A}
+\sin^{4}(\theta/2)D_{B},\nonumber\\
\int_{0}^{\infty }d\tau\langle
B_{33}(0)B_{33}(-\tau)\rangle&=&\int_{0}^{\infty }d\tau\langle
B_{33}(-\tau)B_{33}(0)\rangle^{\ast}\nonumber\\
&=&\sin^{4}(\theta/2)D_{A} +\cos^{4}(\theta/2)D_{B},\nonumber\\
\int_{0}^{\infty }d\tau\langle
B_{22}(-\tau)B_{11}(0)\rangle&=&\int_{0}^{\infty }d\tau\langle
B_{11}(-\tau)B_{22}(0)\rangle\nonumber\\
&=&\cos^{2}(\theta/2)D_{A}^{\ast}
+\sin^{2}(\theta/2)D_{B}^{\ast},\nonumber\\
\int_{0}^{\infty }d\tau\langle
B_{33}(-\tau)B_{11}(0)\rangle&=&\int_{0}^{\infty }d\tau\langle
B_{11}(-\tau)B_{33}(0)\rangle\nonumber\\
&=&\sin^{2}(\theta/2)D_{A}^{\ast}
+\cos^{2}(\theta/2)D_{B}^{\ast},\nonumber\\
\int_{0}^{\infty }d\tau\langle
B_{33}(-\tau)B_{22}(0)\rangle&=&\int_{0}^{\infty }d\tau\langle
B_{22}(-\tau)B_{33}(0)\rangle\nonumber\\
&=&\frac{1}{4}\sin^{2}(\theta/2)\left(D_{A}^{\ast}
+D_{B}^{\ast}\right),
\end{eqnarray}
where the parameters are introduced as
\begin{eqnarray}
D_{A}&=&\int_{0}^{\infty }d\tau \langle
A(0)A(-\tau)\rangle,\hspace{0.2 cm}D_{B}=\int_{0}^{\infty }d\tau
\langle B(0)B(-\tau)\rangle,\nonumber\\
D^{\ast}_{A}&=&\int_{0}^{\infty }d\tau \langle
A(-\tau)A(0)\rangle,\hspace{0.2 cm} D^{\ast}_{B}=\int_{0}^{\infty
}d\tau \langle B(-\tau)B(0)\rangle.\nonumber\\
\end{eqnarray}
Since the noise operators $B_{nn}$ $(n=1,2,3)$ are Hermitian
operators, then we have the relations
\begin{eqnarray}
\int_{0}^{\infty }d\tau\langle
B_{nn}(-\tau)B_{mm}(0)\rangle=\int_{0}^{\infty }d\tau\langle
B_{mm}(0)B_{nn}(-\tau)\rangle^{\ast}.\nonumber\\
\end{eqnarray}
Therefore we can know all the diagonal correlation functions in
Eq.~(\ref{mastereq}) as long as we obtain the expression of $D_{A}$
and $D_{B}$. According to Eqs.~(\ref{aboperators}) and
~(\ref{thermstdensitymatrix}), we can calculate the expression of
$D_{A}$ as follows,
\begin{eqnarray}
D_{A}&=&\sum_{j}g_{1j}^{2}\langle a_{j}^{\dag }a_{j}\rangle
\int_{0}^{\infty }d\tau e^{i\omega _{aj}\tau}\nonumber\\
&&+\sum_{j}g_{1j}^{2}\langle a_{j}a_{j}^{\dag
}\rangle\int_{0}^{\infty }d\tau e^{-i\omega _{aj}\tau }\nonumber\\
&=&\sum_{j}g_{1j}^{2}\langle a_{j}^{\dag }a_{j}\rangle \left[ \pi
\delta ( \omega _{aj})+i\mathbf{P}\frac{1}{\omega
_{aj}}\right]\nonumber\\
&&+\sum_{j}g_{1j}^{2}\langle a_{j}a_{j}^{\dag }\rangle \left[ \pi
\delta( \omega _{aj})
-i\mathbf{P}\frac{1}{\omega _{aj}}\right]\nonumber\\
&=&\textrm{Re}[D_{A}]+i\textrm{Im}[D_{A}],\label{dastarexpression}
\end{eqnarray}
where
\begin{eqnarray}
\textrm{Re}[D_{A}]&=&\lim_{\omega \rightarrow 0^{+}}\pi \varrho
_{1}\left( \omega \right) g_{1}^{2}\left( \omega \right) \left[
2\bar{n}_{1}\left( \omega \right) +1 \right].
\end{eqnarray}
Note that in the third line of Eq.~(\ref{dastarexpression}) we have
used the formula:
\begin{eqnarray}
\int_{0}^{\infty }d\tau e^{\pm i\omega \tau }=\pi \delta \left(
\omega \right) \pm i\mathbf{P}\frac{1}{\omega },
\end{eqnarray}
where the sign ``$\mathbf{P}$" stands for the usual principal part
integral. Similarly, we can obtain the expression of
$\textrm{Re}[D_{B}]$,
\begin{eqnarray}
\textrm{Re}[D_{B}]&=&\lim_{\omega \rightarrow 0^{+}}\pi \varrho
_{2}\left( \omega \right) g_{2}^{2}\left( \omega \right) \left[
2\bar{n}_{2}\left( \omega \right) +1\right].
\end{eqnarray}
Here $\varrho _{1}(\omega)$ and $\varrho _{2}(\omega)$ are the
densities of state of the heat baths of the TLS$1$ and TLS$2$,
respectively. And $\bar{n}_{1}(\omega)=1/[\exp(\beta_{1}\omega)-1]$
and $\bar{n}_{2}(\omega)=1/[\exp(\beta_{2}\omega)-1]$ are the
average thermal excitation numbers. Using the same method we can
obtain the following expressions:
\begin{eqnarray}
&&\textrm{Re}\left[\int_{0}^{\infty }d\tau e^{i\varepsilon \tau
}\langle B_{23}(0)B^{\dag}_{23}(-\tau)\rangle\right]\nonumber\\
&=&\textrm{Re}\left[\int_{0}^{\infty }d\tau e^{-i\varepsilon \tau
}\langle
B_{23}(-\tau)B^{\dag}_{23}(0)\rangle\right]\nonumber\\
&=&\frac{\pi}{4}\sin
^{2}\theta[\varrho_{1}(\varepsilon)g_{1}^{2}(\varepsilon)
(\bar{n}_{1}(\varepsilon)+1)+\varrho_{2}(\varepsilon)g_{2}^{2}(\varepsilon)
(\bar{n}_{2}(\varepsilon)+1)]\nonumber\\
\end{eqnarray}
and
\begin{eqnarray}
&&\textrm{Re}\left[\int_{0}^{\infty }d\tau e^{-i\varepsilon \tau
}\langle B^{\dag}_{23}(0)B_{23}(-\tau)\rangle\right]\nonumber\\
&=&\textrm{Re}\left[\int_{0}^{\infty }d\tau e^{i\varepsilon \tau
}\langle B^{\dag}_{23}(-\tau)B_{23}(0)\rangle\right]\nonumber\\
&=&\frac{\pi}{4}\sin
^{2}\theta[\varrho_{1}(\varepsilon)g_{1}^{2}(\varepsilon)
\bar{n}_{1}(\varepsilon)+\varrho_{2}(\varepsilon)g_{2}^{2}(\varepsilon)
\bar{n}_{2}(\varepsilon)].
\end{eqnarray}
By substituting these correlation functions into
Eq.~(\ref{mastereq}) and returning to the Schr\"{o}dinger picture,
we can obtain quantum master
equation~(\ref{mastereqfordiagonalcase}).


\begin{thebibliography}{99}
\bibitem{Blankenship}  R. E. Blankenship, \emph{Molecular Mechanisms of Photosynthesis} (Blackwell Science, Oxford, 2002).

\bibitem{May-Kuhn}     V. May and O. K\"{u}hn, \emph{Charge and Energy Transfer Dynamics in Molecular Systems} 2nd ed. (Wiley-VCH Verlag, Berlin, 2004).

\bibitem{Fleming1994}  G. R. Fleming and R. van Grondelle, Phys. Today \textbf{47}, 48 (1994).

\bibitem{Fleming2009}  Y. C. Cheng and G. R. Fleming, Annu. Rev. Phys. Chem. \textbf{60}, 241 (2009).

\bibitem{Renger2009}   T. Renger, Photosynth Rev. \textbf{102}, 471 (2009).

\bibitem{Fleming2007}  H. Lee, Y. C. Cheng, and G. R. Fleming, Science \textbf{316}, 1462 (2007).

\bibitem{Flemingnature}G. S. Engel, T. R. Calhoun, E. L. Read, T. K. Ahn, T. Mancal, Y. C. Cheng, R. E. Blankenship,
                       and G, R. Fleming, Nature (London) \textbf{446}, 782 (2007).

\bibitem{Forster1948}  T. F\"{o}rster, Ann. Phys. \textbf{2}, 55 (1948).

\bibitem{Ishizaki2009} A. Ishizaki and G. R. Fleming, J. Chem. Phys. \textbf{130}, 234110 (2009); \textbf{130}, 234111 (2009).

\bibitem{Jang2008}     S. Jang, Y. C. Cheng, D. R. Reichman, and J. D. Eaves, J. Chem. Phys. \textbf{129}, 101104 (2008).

\bibitem{Palmieri2009} B. Palmieri, D. Abramavicius, and S. Mukamel, J. Chem. Phys. \textbf{130}, 204512 (2009).


\bibitem{Aspuru-Guzik2008}  M. Mohseni, P. Rebentrost, S. Lloyd, and A. Aspuru-Guzik, J. Chem. Phys. \textbf{129}, 174106 (2008).

\bibitem{Aspuru-Guzik20091} P. Rebentrost, M. Mohseni, I. Kassal, S. Lloyd, and A. Aspuru-Guzik, New J. Phys. \textbf{11}, 033003 (2009).

\bibitem{Aspuru-Guzik20092} P. Rebentrost, M. Mohseni, and A. Aspuru-Guzik, J. Phys. Chem. B \textbf{113}, 9942 (2009).

\bibitem{Aspuru-Guzik20093} P. Rebentrost, R. Chakraborty, and A. Aspuru-Guzik. J. Chem. Phys. \textbf{131}, 184102 (2009).
\bibitem{Plenio2008}   M. B. Plenio and S. F. Huelga, New J. Phys. \textbf{10}, 113019 (2008).

\bibitem{Plenio20091}  F. Caruso, A. W. Chin, A. Datta, S. F. Huelga, and M. B. Plenio, J. Chem. Phys. \textbf{131}, 105106 (2009);
\bibitem{Castro2008}   A. Olaya-Castro, C. F. Lee, F. F. Olsen, and N. F. Johnson, Phys. Rev. B \textbf{78}, 085115 (2008).

\bibitem{Castro2009}   F. Fassioli, A. Nazir, and A. Olaya-Castro, J. Phys. Chem. Lett. \textbf{1}, 2139 (2010).

\bibitem{Nazir2009}    A. Nazir, Phys. Rev. Lett. \textbf{103}, 146404 (2009).

\bibitem{Nori2009}     A. Yu. Smirnov, L. G. Mourokh, and F. Nori, J. Chem. Phys. \textbf{130}, 235105 (2009);
                       P. K. Ghosh, A. Yu. Smirnov, and F. Nori, J. Chem. Phys. \textbf{131}, 035102 (2009);
                       A. Yu. Smirnov, S. Savel'ev, and F. Nori, Phys. Rev. E \textbf{80}, 011916 (2009);
                       A. Yu. Smirnov, L. G. Mourokh, P. K. Ghosh, and F. Nori, J. Phys. Chem. C. \textbf{113}, 21218 (2009).

\bibitem{Liang2010}    X. T. Liang, W. M. Zhang, and Y. Z. Zhuo, Phys. Rev. E \textbf{81}, 011906 (2010).

\bibitem{Yang2010}     S. Yang, D. Z. Xu, Z. Song, and C. P. Sun, J. Chem. Phys. \textbf{132}, 234501 (2010).

\bibitem{Bell1987}     S. J. Bell, \emph{Speakable and Unspeakable in Quantum Mechanics} (Cambridge University Press, Cambridge, 1987).

\bibitem{Einstein1935} A. Einstein, B. Podolsky, and N. Rosen, Phys. Rev. \textbf{41}, 777 (1935).

\bibitem{Nielsen2000}  M. A. Nielsen and I. L. Chuang, \emph{Quantum Computation and Quantum Information} (Cambridge University Press, Cambridge, 2000).

\bibitem{Qian2005}     X. F. Qian, Y. Li, Y. Li, Z. Song, and C. P. Sun, Phys. Rev. A \textbf{72}, 062329 (2005).

\bibitem{Briegel2008}  H. J. Briegel and S. Popescu, arXiv:0806.4552.

\bibitem{Briegel2009}  J. Cai, G. G. Guerreschi, and H. J. Briegel, Phys. Rev. Lett. \textbf{104}, 220502 (2010).

\bibitem{Thorwart2009} M. Thorwart, J. Eckel, J. H. Reina, P. Nalbach, and S. Weiss, Chem. Phys. Lett. \textbf{478}, 234 (2009).

\bibitem{Sarovar2009}  M. Sarovar, A. Ishizaki, G. R. Fleming, and K. B. Whaley, Nat. Physics \textbf{6}, 462 (2010).

\bibitem{Caruso2009}   F. Caruso, A. W. Chin, A. Datta, S. F. Huelga, and M. B. Plenio, Phys. Rev. A \textbf{81}, 062346 (2010).

\bibitem{Leggettnp}    A. O. Caldeira and A. J. Leggett, Ann. Phys. (N.Y.) \textbf{149}, 374 (1983).

\bibitem{Gao2007}      Y. B. Gao and C. P. Sun, Phys. Rev. E \textbf{75}, 011105 (2007).

\bibitem{Breuer}       H. P. Breuer and F. Petruccione, \emph{The Theory of Open Quantum Systems} (Oxford University Press, Oxford, 2002).

\bibitem{Wootters1998} W. K. Wootters, Phys. Rev. Lett. \textbf{80}, 2245 (1998).

\bibitem{Zubairy1998}  M. Ikram, F. L. Li, and M. S. Zubairy, Phys. Rev. A \textbf{75}, 062336 (2007).

\bibitem{YuTing}       T. Yu and J. H. Eberly, Phys. Rev. Lett. \textbf{93}, 140404 (2004); \textbf{97}, 140403 (2006);
                       Phys. Rev. B \textbf{68}, 165322 (2003); Science \textbf{323}, 598 (2009).

\bibitem{Almeida}      M. P. Almeida, F. de Melo, M. Hor-Meyll, A. Salles, S. P. Walborn, P. H. Souto Ribeiro, and L. Davidovich, Science \textbf{316}, 579 (2007).

\bibitem{Ficek}        Z. Ficek and R. Tana\'{s}, Phys. Rev. A \textbf{74}, 024304 (2006).

\bibitem{FQWang}       F. Q. Wang, Z. M. Zhang, and R. S. Liang, Phys. Rev. A \textbf{78}, 062318 (2008).

\bibitem{Dubi}         Y. Dubi and M. Di Ventra, Phys. Rev. A \textbf{79}, 012328 (2009).

\bibitem{BAn}          N. B. An and J. Kim,  Phys. Rev. A \textbf{79}, 022303 (2009).

\bibitem{Ann}          K. Ann and G. Jaeger, Phys. Rev. B \textbf{75}, 115307 (2007).

\bibitem{ZheSun}       Z. Sun, X. G. Wang, and C. P. Sun, Phys. Rev. A \textbf{75}, 062312 (2007).

\bibitem{HZheng}       X. F. Cao and H. Zheng, Phys. Rev. A \textbf{77}, 022320 (2008).

\bibitem{James}        A. Al-Qasimi and D. F. V. James, Phys. Rev. A \textbf{77}, 012117 (2008).

\bibitem{Bellomo}      B. Bellomo, R. Lo Franco, and G. Compagno, Phys. Rev. Lett. \textbf{99}, 160502 (2007).

\bibitem{Davidovich}   L. Aolita, R. Chaves, D. Cavalcanti, A. Ac\'{i}n, and L. Davidovich, Phys. Rev. Lett. \textbf{100}, 080501 (2008).

\bibitem{Ban}          M. Ban, Phys. Rev. A \textbf{80}, 032114 (2009).

\bibitem{Scholak}      T. Scholak, F. de Melo, T. Wellens, F. Mintert, and A. Buchleitner,  arXiv:0912.3560.
\end{thebibliography}
\end{document}